\documentstyle[11pt,aaspp4]{article}
\def\etal{{et al.~}}

\begin{document}
\noindent

\vskip .2in
\baselineskip 20pt
\singlespace

\title{The Burst and Transient Source Experiment\\
Earth Occultation Technique}
\title{Short title: BATSE Earth Occultation}
\centerline{B. A. Harmon\footnote{alan.harmon@msfc.nasa.gov}, G. J. Fishman,
\& C. A. Wilson}
\centerline{NASA Marshall Space Flight Center, 
SD50, Huntsville, AL 35812 USA} 
\vskip .2in
\centerline{W. S. Paciesas\footnote{bill.paciesas@msfc.nasa.gov}\& S. N.
Zhang}
\centerline{Department of Physics,
University of Alabama in Huntsville, Huntsville, AL USA 35899}
\vskip .2in
\centerline{M. H. Finger\footnote{mark.finger@msfc.nasa.gov}, T. M. Koshut,
M. L. McCollough, 
C. R. Robinson\footnote{current address:National Science Foundation,
Arlington, VA 22230; crobinso@nsf.gov}
\& B. C. Rubin\footnote{current address:American Physical Society, Ridge,
NY, 11961; rubin@aps.org}}
\centerline{Universities Space Research Association, SD50, 
NASA Marshall Space Flight Center, Huntsville, AL 35812 USA}

\begin{abstract}

An Earth orbiting detector sensitive to gamma ray 
photons will see step-like occultation features in its counting rate
when a gamma ray point source crosses the Earth's limb.  
This is due to the change in
atmospheric attenuation of the gamma rays along
the line of sight.
In an uncollimated detector, these occultation features can be used
to locate and monitor
astrophysical sources provided their signals can be individually
separated from
the detector background. We show that the {\it Earth occultation
technique} applied to the Burst and Transient Source Experiment
(BATSE) on the {\it
Compton Gamma Ray Observatory} (CGRO) is a viable and flexible
all-sky monitor in the low energy gamma ray and hard X-ray energy
range (20 keV - 1 MeV).  The method is an alternative to more sophisticated
photon imaging devices for astronomy, and can serve well as a cost-effective 
science capability for monitoring the high energy sky.

Here we describe the Earth occultation technique for locating new sources
and 
for measuring source intensity and spectra without the use of complex 
background models.  Examples of transform imaging, step searches,
spectra, and light curves are presented.  Systematic uncertainties due to
source confusion, detector response, and contamination from rapid background
fluctuations are discussed and analyzed for their effect on intensity
measurements.  A sky location-dependent average systematic error is derived
as a function of galactic coordinates.  The sensitivity of the technique
is derived as a function of incident photon energy and also as a function
of angle between the source and the normal to the detector entrance window.
Occultations of the Crab Nebula by the Moon are used to calibrate 
Earth occultation flux measurements independent of possible atmospheric
scattering effects.

\end{abstract}


\newpage

\section{INTRODUCTION}

The {\it Compton Gamma Ray Observatory} (CGRO), one of
NASA's four Great Observatories, was launched
in 1991 April and operated in
low Earth orbit until controlled re-entry
in 2000 June.  CGRO was 
responsible for many discoveries in the study of gamma ray bursts,
accreting binaries, active galaxies, and pulsars (Gehrels \& Shrader 1997;
Kniffen \& Gehrels 1997; Leonard \& Wanjek 2000).  The quest for the origin of gamma ray bursts
led to the development and flight of The Burst and Transient Source
Experiment (BATSE) on
the CGRO. BATSE
pointed the way to the extragalactic origin of gamma ray bursts
through mapping of burst location and number-brightness distributions
(Meegan \etal 1992).
In addition to BATSE's primary science goals, its nine years of nearly
continuous operation
and all-sky capability allowed monitoring of the low energy gamma ray/hard
X-ray sky using the
Earth occultation technique (EOT).  Prior to BATSE, the method had not
been used widely by the astrophysics community for gathering spectral and
intensity information about celestial sources.  We show that the Earth
occultation technique, applied to a well-calibrated multiple detector system
and without sophisticated imaging hardware, is a viable and inexpensive
method for monitoring the high energy sky.  It is particularly suitable
for uncollimated devices such as background or anti-Compton suppression
shields, which are commonly used in X-ray and gamma-ray astronomy missions.  
We therefore present here a detailed discussion of how the Earth occultation 
technique was applied to BATSE data.  A listing of acronyms and abbreviations used 
throughout the text can be found in Appendix A.
 
Historically, monitoring of the high energy sky was accomplished in the energy 
range of $\sim$1-20 keV using various types of scanning pinhole cameras, 
modulation collimators, or position-sensitive detectors.
X-ray missions such as {\it Uhuru} (Giacconi
\etal 1972), {\it Vela 5B} 
(Conner, Evans \& Belian 1969), the {\it Ginga}
All-Sky Monitor (Tsunemi \etal 1989), 
{\it Ariel 5} (Holt 1976) the GRANAT and EURECA Watch monitors (Lund, 1986;
Brandt, 1994; Castro-Tirado, 1994)
and most recently, the All-Sky Monitor on the {\it Rossi X-Ray Timing
Explorer} (RXTE) (Bradt, Rothschild \& Swank 1993; Levine \etal 1996) have 
been crucial in discovering and 
monitoring of
X-ray transients, investigating long-term periodic variations in more
persistent sources, and detecting other phenomena associated with the 
X-ray sky.  BATSE complements these instruments by monitoring the energy
range
from 20 keV up to about 1 MeV.  This covers the lower energy portion of the 
nonthermal regime, where
the emission is produced by a variety of processes such as Compton
up-scattering
of soft photons by energetic particles, bremsstrahlung, and synchrotron
radiation.

It was realized prior to launch (Fishman \etal 1982; 1984; 1989) that BATSE
could
serve as a sensitive all-sky monitor of point-like 
gamma ray sources. Earth occultations of point sources allow sampling
of high energy fluxes using the sharp step-like features in
the background data for both pulsed and nonpulsed sources (Paciesas \etal
1985). 
Variations on timescales of hours to years can be investigated.  
The method allowed BATSE to be very successful at
locating new transients, detecting unusual intensity or spectral changes,
and stimulating observations in other wavelength
bands.  Because the EOT was enhanced and used extensively during
the operational period of CGRO,
we present its use in various forms, and discuss
a number of aspects of the
technique such as source identification, sensitivity, and systematic error.
Pre-launch discussions of the method can be
found in Paciesas \etal (1985) and Fishman \etal (1989). Discussions 
concerning the
post-launch performance and expanded capabilities can be found in Harmon
\etal
(1992), Wilson \etal (1992), Zhang \etal (1993), and Zhang \etal (1994a). 
Preliminary
surveys of detected sources can be found in Harmon \etal (1993) and Robinson
\etal (1997).   An imaging capability was added after launch, based on the
Radon
transform, which greatly improved our ability to locate
and identify gamma ray sources (Zhang \etal 1993, 1994a).  Another powerful 
application of BATSE as an all-sky monitor was the ability to
to detect and monitor pulsars using Fourier
analysis or epoch folding techniques. A detailed description of these
methods and results from BATSE data can be found in Bildsten \etal (1997). 

\section{INSTRUMENTATION AND OCCULTATION GEOMETRY}

\subsection{BATSE Instrumentation}

The Earth occultation analysis technique 
utilizes the large area detectors (LADs) on BATSE, which are
sensitive to photons above 20 keV.  The {\it Compton Gamma Ray Observatory}
and placement
of the BATSE detector modules on the spacecraft are shown in Fig. 1.
The LADs are composed of sodium iodide (NaI(Tl)) crystals, 1.27 cm (0.5 in) thick by
50.8 cm (20 in) across (2025 cm$^{2}$ total area of one detector).  Eight 
modules 
are mounted on
the corners of the CGRO with normal vectors perpendicular to the faces
of a regular octahedron.  Any point on the
sky can viewed by four detectors at angles less than 90$^{\circ}$ to
the source direction.  This gives BATSE a
capability of obtaining crude locations to within a few  degrees using the 
count rates from a combination of detectors and the known response
of each detector from photons directly from the source, and source photons
scattered off
the atmosphere (Pendleton \etal 1995a).  This is routinely done to 
locate gamma ray bursts to within an accuracy of a few degrees (Pendleton \etal 1999).   
A full description of 
 the BATSE detectors can be found in a number of places; 
see, for example, Fishman \etal (1984, 1989).

Although Earth occultation features have been used with other
gamma ray instruments to restrict
the location of point sources (e.g., Wheaton \etal 1982), to our knowledge,
only BATSE made use of the method
for direct measurement of point source emission.
Some Earth occultation measurements have been performed
with the BATSE spectroscopy detectors (SDs) (McNamara, Harmon \& Harrison
1995;
1996; Paciesas et al. 1998; McNamara \etal 1998).  These detectors have a
much smaller effective area for source monitoring ($\sim$600 cm$^{2}$) than the LADs, but
do have an additional low energy bandpass around 10 keV.  Here we 
discuss the method as applied only to the large area detectors.  

The Earth occultation technique was also developed and applied to BATSE data in
a separate, parallel effort at the Jet Propulsion Laboratory (JPL) (Skelton
\etal 1994; Ling \etal 1996).  The JPL group has recently published a
description of their approach (the Enhanced BATSE Occultation Package (EBOP)), 
and a compendium of measurements using the technique covering the 1991 May to
1994 October epoch (Ling \etal 2000).  The method used by JPL was developed
specifically for extraction of flux histories and spectra of point gamma ray
sources.  There are primarily two major differences between
the JPL method and the method we describe in this paper. First, EBOP uses a 
semi-physical model for the detector background counting
rates. The model is based on expected contributions of low energy gamma ray fluxes
local to the low Earth orbit (LEO) environment.  These include cosmic ray
secondary radiation and activation products from orbital passes through
radiation fields in LEO.   The JPL 
global background model consists of a mix of
the local radiation components and a combination (determined by the fit)
of the SD counting rates as a predictor of the low energy background in the
LADs.   Secondly, the extraction of source signals is performed usually in 
one-day segments,
with a single fit including terms for {\it all} sources in the EBOP catalog
with no{\it a priori} assumptions of their intensity. 
The JPL method has the advantage of potentially greater statistical
accuracy at the expense of increased systematic error.  The method appears to
give reasonable results for bright, hard sources such as the Crab Nebula
and Cygnus X-1.  In some cases, for relatively weak sources, 
such as the neutron star binaries Circinus X-1 and Scutum X-1, the JPL method 
yields significant hard emission greater than $\sim$200 keV, which is
uncharacteristic for this class of sources (see discussion 
in Sec. 3.2 and Fig. 8, plates
10 and 75 of Ling \etal 2000).  We have not been able to confirm
these unusually hard spectra with the 
Marshall Space Flight Center (MSFC) method, nor are they reported by other
high energy observatories.  

Instead of the physical inputs used in the JPL fitting
procedure, the MSFC method uses simple polynomials as the set of basis 
functions for fitting the background and extracting the source signal 
(see Section 3 for details). This is equivalent to assuming that the background
is smooth 
on a timescale of a few minutes, as discussed in Section 3, with
respect to count rate variations caused by cosmic ray secondaries, activation, 
and other local background
components.  This assumption breaks down only during times of high solar flare
activity, gamma ray bursts and a few flaring gamma ray sources, which are 
either excluded from the analysis,
or flagged later at the discretion of the observer (see Sec. 2.2 for a
discussion of data selection procedures). The only source terms included in the
fit are those which are thought to be active during the short four-minute time
windows.  We chose to adopt the method described here over that of EBOP 
to minimize the effect of unpredictable systematic error, and because of
the real-time needs of the MSFC all-sky monitoring effort.  The
MSFC method also lends itself well to an iterative approach in selecting sources
that may potentially interfere with the source of interest (SOI), so that a best
solution comes from building up a knowledge of the sky region within a few degrees
surrounding the source.   

More recently, Southampton University has undertaken a more rigorous approach
in modeling the
BATSE backgrounds in order to generate all-sky images and extract of
source fluxes using Earth occultation (Shaw \etal 2000).  This method
incorporates the physical background components discussed above and a mass model of the {\it Compton
Observatory} in a Monte Carlo simulation using the radiation transport
code GEANT as was done with the European Space Agency International Gamma
Ray Astrophysics Laboratory (INTEGRAL) (Lei \etal 1999).  The result of the simulation is the {\it a priori}
determination of the total diffuse background in the LADs at any orientation and geomagnetic
coordinates of the Earth.  This methods offers the possibility of removing
the source signal without the need for a simultaneous fit to the background.

\subsection{Data Selection}

The LADs are well-suited to Earth occultation measurements because
of their sensitivity, uniformity in energy range, and stabilized gain.  
Data from two or more LADs can be easily
combined or fitted jointly, depending on the application. 
There are two different data types which are
most convenient for occultation measurements: DISCLA (LAD
discriminator data) and CONT (or LAD continuous data) which cover
the same energy range (20 keV - 1 MeV).  The DISCLA data type
provides four energy channels binned every 1.024 seconds,
and the CONT datatype provides 16 channels every 2.048 seconds.

Prior to application of the Earth occultation method,
proper data selection is required to remove large fluctuations that may
affect the fitting of occultation steps.  When data are flagged in one stage
of the
selection process, they are then excluded 
from the occultation analysis. 

The first stage of the data selection is performed onboard.  Several
different datatypes
are scanned, including the DISCLA type, by comparing the local background
count rate to new data as they are acquired.  An event may ``trigger" the
gamma ray
burst (GRB) acquisition mode when a high data rate is encountered relative
to the
background in two or more LADs (Meegan \etal 1992).   High rate events may
be caused by bursts,
solar flares, 
bremsstrahlung from electron precipitation (Aschwanden, Schwartz \&
Dennis 1998),
terrestrial gamma ray flashes (Fishman \etal 1994a), or even flaring
galactic
sources (Mallozzi \etal 1993). Any triggered events are flagged
automatically in the datastream
and identified later. 

The second stage of the selection process consists of manual inspection
for other events that may escape the burst trigger, yet generate large 
transient flares.  These are of a somewhat
longer timescale than GRBs, usually a
few tens of seconds to minutes.  The fluxes from these
events are generally easy 
to identify based on their observed timing and spectral properties.
Time periods
containing transient background features large enough to be detected 
visually (about 10\% of the total background
counting rate) are flagged and excluded from occultation analysis along
with the triggered events.  The vast majority of these events are
gamma ray fluxes solar flares or bremsstrahlung
from precipitating electrons.

The last stage of the data selection consists of
additional flagging of very short ($<$ 1 sec) cosmic ray events. These occur 
in only one LAD at a time when a heavy cosmic ray, such as an iron nucleus,
deposits a large amount of energy into the scintillation medium. This is 
observed as a sharp, positive-going spike in the background data for one time 
bin.  It is caused by long-lived phosphorescence characteristic of impurities 
in the NaI(Tl) crystal (Fishman \& Austin 1976).  The affected data bins are 
flagged using a spike filter prior to occultation as well as pulsar 
measurements.

An interesting aside to the standard data treatment discussed above is that
bright and extremely variable galactic sources
are occasionally detected via the burst trigger.  Such occurrences
do not affect earth occultation analyses to a large degree because
the amount of data excluded in the data selection process due to intrinsic
source variability is statistically insignificant.
However, the extremes of variability are reduced if
the burst trigger is enabled by the SOI, since the CONT
or DISCLA data at the time of the trigger
are rejected in our data analysis procedure.  The sources for which this was
known to have occurred during the mission
are A 0535+262 (Finger, Wilson
\& Harmon 1996) (see Fig. 2), Cygnus X-1 (Fishman
et al. 1994b), GRO J0422+32 (Mallozzi \etal 1993), 
GRO J1744-28 (Kouveliotou \etal 1996), and 4U 1700-377
(Rubin \etal 1996a), and soft gamma ray repeaters (Woods \etal 1999). 
For some flaring episodes of these sources, the BATSE burst trigger
threshold
was raised temporarily in order to minimize the number of non-GRB
triggers.
Normally, the likelihood of a source being rejected in this manner for
Earth occultation
measurement is rather small due to the great difference in peak brightness
of these sources and that of GRBs.  It should be noted that when
data are flagged and removed from the analysis, 
in actuality, the data are recoverable by re-analyzing the CONT or DISCLA data
with less restrictive filtering criteria.

\subsection{Source Occultations}

In one orbit around the Earth,
two occultation step features, a rise and set pair, will be superimposed on
the background counting rate as each point source is occulted.  
A measurement can be made of the intensity of a source in each energy
channel at rise or set.  In practice,
two measurements per {\it every} orbit are not achieved.  The most common
reasons
are passages through the lower Van Allen radiation belt at the
South Atlantic Anomaly when the detector voltage is turned off,
or that CGRO is out of line-of-sight contact with the NASA Tracking \& Data
Relay Satellites (TDRSS),
and data which have been flagged and are not available for analysis. High
declination
sources ($\sim|\delta|\geq$41$^{\circ}$) also experience
an interruption of occultations near the orbital poles (see Appendix B).  Source confusion,
where occultations of one
source are indistinguishable from another, also limits the number
of usable occultation steps.  The impact of these effects combined causes 
Earth occultation coverage averaged over one precession cycle ($\sim$52d)
to range between 80\%-90\%, at best, and at worst, about 50\%. 

The attenuation of gamma rays by the Earth's atmosphere and the
variation in thickness of the air mass along the line of sight to the
X-ray source produce the step-like features in the detector count rate as
a function of time.  The attenuation is 50\% for 100 keV photons that 
pass through the atmosphere along a line of sight with minimum altitude of
70 km.
For a typical orbital speed of 8 km/s, the duration
of the occultation step for a source rising or 
setting in the plane of the spacecraft orbit is about 10 seconds.
Therefore occultations
are relatively sharp features superimposed
on the generally slower background variations caused
by orbital motion around the Earth.  

Several examples of occultation features, or ``steps", in the DISCLA 
and CONT data are shown in Fig. 2.  Except where otherwise noted, the date
and time convention used in this paper is {\it Truncated Julian Date} (TJD)
= Julian Date
(JD) - 2,440,000.5 and {\it seconds of day} measured from the TJD start. 
Fig. 3 shows a close-up view
of individual occultation steps with fitting functions based on a model for
the
atmospheric attenuation for gamma rays (See Section 3.1).

The accuracy to which the timing and magnitude of these steps can be fit
determines
the performance of the Earth occultation technique.  We discuss three forms 
of the technique from which information can be obtained using BATSE data:  
(1) the flux and spectra extraction method, where we assume {\it a priori}
knowledge of source locations, and (2) the step search and (3) the
occultation imaging methods,
where the source location is not required.  Methods (2) and (3) are
best-suited for
new source searches. An overall comparison of these approaches is given in
Table 1.

In Appendices B and C, we supplement the discussion of the EOT
by developing the mathematical framework for Earth occultation from an
orbiting spacecraft. This includes expressions for timing of Earth occultation
features and their use for locating point sources.  

\section{FLUX AND SPECTRA EXTRACTION FOR KNOWN SOURCES}

\subsection{Estimation of Occultation Steps in Counting Rates}

The count rate for a source in the LAD is extracted by
simultaneously fitting occultation step features with terms
for each source in the fit and a quadratic polynomial to represent
the detector background.  The fit is performed independently for each
energy channel.  
Each occultation step, rise or set,
is fit over a time {\it t$_{occ}$$-\tau$ to \it t$_{occ}$+$\tau$}, 
which we refer to as the {\it fitting window}.
{\it t$_{occ}$} represents the occultation time of the 
SOI and also the center of the time window.  Use of a
quadratic
form for the background restricts the half-width of the fitting window,
$\tau$, to no more than
about 120 seconds of data.
The modeled detector count rate $R$ in each energy channel can be
represented as
\begin{equation}
R(t) = \sum_{i=0}^{2} b_{i} (t-t_{occ})^{i} 
                               + \sum_{j=1}^{n} r_{j}T_{j}(t)
\label{eqn:mod_rate}
\end{equation}
where $b_i$ are the coefficients of the background model (to second order) 
and $r_j$ are the source count rates including the SOI and other
bright sources with occultation steps occurring within the fitting window, 
and $T_{j}(t)$ are the atmospheric transmission functions. The number of
source terms {\it n} is kept to a minimum for greatest
sensitivity, at the expense of incurring some systematic error.  This method
is 
similar to that
adopted in Wheaton \etal (1995) for multi-parameter least squares fitting of
data
when the detector background is time variable and individual measurements
are of low
statistical quality.  We assume that the background is smooth and adequately
fit by the second order polynomial on the order of the size of the fitting
window, 2$\tau$.  In particular, the failure of this assumption is usually
caused by the presence in the data of bright pulsars, weak bursts,
solar flares, and other disturbances on the timescale of the fitting window. 
Problematic stretches of data are mostly removed in the data selection
process discussed in Sec. 2.2.  Experience shows that there are non-Poissonian
components that remain in the data, and must be accounted for in the analysis
of results.  However, most of these effects add incoherently on timescales of
a day or longer.  

The occultation features are represented 
via a model function for the transmission $T(t)$ as
\begin{equation}
T(t) = e^{\mu(E)A(h(t))}
\label{eqn:transm}
\end{equation}
where $\mu(E)$ is the energy-dependent mass attenuation coefficient of gamma
rays in air (Storm \& Israel 1970; Chupp 1975) and {\it A(h)} is the air
mass 
computed along the line of sight
at a given altitude {\it h(t)}.  {\it A(h)} is interpolated from a table of 
air masses (W. Wheaton, private communication) for values of {\it h} between
50 and 110 km, and is
based on the U.S. International Commercial Aviation Organization (ICAO) Standard
Atmosphere (1962). 
Use of Eqs.~(\ref{eqn:mod_rate}) and (\ref{eqn:transm}) requires precise
knowledge of the spacecraft ephemeris (time and position), the 
direction to the SOI, 
and a model of the Earth that accounts for its non-sphericity (see
Appendix C).  To use the CONT data at its full time resolution of 2.048s, the position
of the spacecraft at the center of the time bin must be known accurately.
The spacecraft position is
nominally interpolated for the center of the 2.048s time bin from the incoming 
ephemeris data.
For most of CGRO mission, the component of the position vector in
the direction of spacecraft motion was known
to less than 10 km.  Due to the sharpness of occultation profiles ($\sim$10s),
ephemeris errors even on the order of
25 km downrange (about 1 part in 2000) can shift the step model enough
to seriously affect the flux measurement.  A few ephemeris problems occurred
during the mission, but were corrected in the data archive.
The ability to measure the
time of an occultation and the sharpness of Earth occultation features 
can be exploited to determine the location of a source as 
outlined in Appendices B and C. 

A database of source locations, outburst times, and intensities
provides information to determine whether
terms for sources should be included in the fitting window.  This database
was built up as new
sources were found either through occultation analysis (light curves and
images) or from other instrument measurements.  Information about source
outburst
intensity levels as a function of time are read from the database by the
analysis software before the flux measurement is performed. 

Each energy channel in the source-pointed LADs (defined to be
less than 60$^{\circ}$ between
the source direction and detector normal) is fit independently to 
derive a count rate in each CONT channel for all sources {\it r$_{j}$}
as a function of time. The statistical error ($\delta r_{j}$) of the $r_{j}$th term at 
time $t_{occ}$ is computed from the least squares
fit of Eq.~(\ref{eqn:mod_rate}).  Some physical insight into the obtainable
error from the Earth occultation fits can be achieved
by assuming a simple linear
step function in place of the transmission function $T_{j}(t)$ in  
Eq.~(\ref{eqn:mod_rate}) and setting $j$=1 (only the SOI is considered).   
The uncertainty in the fit can then be extracted from the Hessian, or information, matrix
(Press \etal 1992), which is to first order  
\begin{equation}
\delta r_{j} = \eta \sqrt{\frac{2R_{o}}{\tau}}
\label{eqn:rterr}
\end{equation}
where $\eta$ is a parameter that depends on the half-width, $\tau$, the occultation step width, 
and the slope of the background across the
fitting window. $R_{o}$ is the detector count rate at the
center of the fitting window.  For an occultation step of width
10s and 2$\tau$=240s, $\eta$$\approx$3.  
  
To derive a flux history, the fitting coefficient
corresponding to the SOI (the $r_{j}$th
term at time $t_{occ}$ in Eq.~(\ref{eqn:mod_rate}))
is accumulated as a function of time for later deconvolution from the instrument
response.  However, all the coefficients,
including those for other sources in the fit, and the background
terms, can be written to a file for later use.   

\subsection{Spectral Analysis}

Here we present examples illustrating the 
usefulness of the BATSE EOT to measure source
variability and 
differences in spectral behavior. The technique has been used
to monitor 
spectral state transitions in the black hole candidates 
GX 339-4 (Harmon \etal 1994; Rubin \etal 1998) and Cyg X-1 (Zhang \etal
1997).  During the operational lifetime of CGRO, outbursts 
of monitored sources, particularly unusual ones, were made known to the
scientific
community through various electronic media.

The history of the source intensity and spectral behavior can be generated from the 
{\it r$_{j}$} in Eq.~(\ref{eqn:mod_rate}).  The {\it r$_{j}$} in counts
s$^{-1}$ per energy channel per LAD, 
without correction for the detector response, are stored in a 
time-ordered file with a beginning and ending time specified
by the user.   Energy spectra (flux per energy channel) and light curves showing
the history
of the source's intensity
as a function of time can be generated from the raw history file.  

The octahedral geometry of BATSE insures that at least four detectors
simultaneously view a point source anywhere on the sky.  The angular sensitivity
of the LADs is maintained by combining statistics from two, three and ultimately
four
detectors at successively larger angles from any one of the detector normals as
discussed later in Sec. 3.3.  Since there is no improvement in the signal-to-noise ratio by 
combining data from detectors beyond about 60$^{\circ}$, this is a convenient 
cutoff angle for choosing which detectors are used. Furthermore,
the natural timescale for which spectral measurements are most easily obtained
without changing detector combinations is the
observation, or pointing period, of CGRO.  Observations, where the orientation
of the LADs were kept fixed with respect to celestial coordinates,
lasts about two to three weeks.   The specific orientation of the X and
Z axes, which were set by the CGRO observing schedule, determined the
LAD combination for viewing a point on the sky.  

In Fig. 4, LAD count spectra and model residuals
are shown for four 
sources extracted using the method of the previous section. Each
frame consists of source count rates as measured via Earth occultation for
a set of LADs whose normal vectors are less than 60$^{\circ}$ to the
respective direction of the source.  All occultations of same source have
been fitted to determine the {\it r$_{j}$}, which are then averaged
over the observation period to yield the rates (error-weighted) and uncertainty
for each channel in each detector as shown in Fig. 4. 
The number of
detectors per combination ranges from two (Vela X-1) to four (Cygnus X-1).  

The dotted curves represent the best fitting model spectra for each LAD.
The fitting procedure, used commonly in high energy astronomy, is called 
``forward-folding" (Briggs 1995).   This method 
avoids a potentially unstable inversion of the BATSE response matrices, 
which have large off-diagonal elements due to the LAD's relatively shallow detection
depth.  (We will
delay discussion of the LAD response formalism until Sec. 3.5.)
Minimization of the chi-square statistic is performed using the 
Levenberg-Marquardt formulation of linear least squares (Press \etal 1992).  
The selected spectral model is folded through the instrument 
response (Pendleton \etal
1995a) for all detectors in the combination to determine model count 
spectra.  
The choice of an appropriate spectral model is
based on trial and error, and depends on the type of source.
The parameters of the model are then
adjusted to minimize chi-square 
between the model count rates and the observed count rates for
all detectors in the combination.  If all fourteen energy channels are used,
the number of degrees of freedom is (\# of LADS) $\times$ 14 - {\it p}, where
{\it p} is the number of parameters in the model.  
Residuals from the model fit in number of sigmas
are shown below the count rate spectra for all detectors in the fit.
 
In Fig. 5 we show the corresponding multiple detector photon
spectra after deconvolution of the response for the same four sources 
and observation periods as in Fig. 4.
The solid curves represent the best fitting spectral model with parameters
given in Fig. 5.  Uncertainties in the fit parameters are obtained according to 
the prescription of
Lampton, Margon \& Bowyer (1976).
Data points in photon space and associated uncertainties (here in 
photons cm$^{-2}$s$^{-1}$keV$^{-1}$) are
derived by multiplying the spectral model by the ratio of the model counts
to the observed counts.   The channel boundaries and photon data points
are averaged for the LADs after deconvolution of the response function.

Representative light curves as a function of time are shown in Fig. 6.
For generation of light curves, the same type of spectral fit is
performed on the count spectra, usually with one or more days of occultation
data, then fluxes for
each point in the light curves
are obtained by integrating over a specified energy range of the best
fitting
spectral model. 
In cases where the flux is not a parameter, the uncertainty in the flux
is obtained either by analytically or numerically differentiating
the fitting model with respect to the parameters.    
For generating light curves, the simpler spectral models, with one or two parameters, such as 
single power law or bremsstrahlung are
best for
the shorter energy range, whereas
for longer integrations as shown in Fig. 5, more sophisticated models 
can be used.

It should be kept in mind that the treatment of statistical errors in the data shown in Figs.
4-6 are based on propagating the error from the source count rates
extracted from occultation
step fits to the raw data.  The errors shown in the count spectra in Fig. 4
come from direct error-weighted averaging of occultation
step rates occurring over an observation period. In Figs. 5 and 6, even 
though the proper
prescription of errors in the spectral fitting parameters are given, it
is important to note that the ``data" in photon space are model dependent. 
For example, when a source is near the minimum detectable limit (see Sec. 3.3), 
and the statistical significance is low, the absolute value of the data
in photon space depends heavily on the choice of spectral model.  A much more
desirable method of obtaining accurate fluxes and/or upper limits in a given 
energy band is to compare the source count rate to that of the Crab Nebula.       

Error-weighted averaging across CGRO pointing boundaries
{\it after} deconvolution
of the response is a straightforward method to increase the significance of 
a source signal.  This is desirable for cases in searching for weaker emission
or longer-term variations in sources.
For the light curves in Fig. 6, it is simply a matter of
binning the flux data into longer time bins.    Alternatively, it is possible to
combine 
channel by channel data {\it before} deconvolution of the response.  The
fitting can then be done by retaining the total exposure to the source in each
observation interval and weighting the detector response accordingly. This 
approach is desirable when energy spectra for weak sources ($\sim$few mCrab) 
are required;
however, systematic errors limit the ultimate flux sensitivity that can be achieved
by averaging over many weeks or even years.  The issue of systematic error
is dealt with in later sections.

\subsection{Sensitivity}

We first consider the sensitivity of Earth occultation without complicating
factors such as nearby point sources and imperfections in the detector
response function. 
The sensitivity of the EOT with the BATSE LADs
depends on several factors.  The uncollimated detector geometry of 
the LADs and
the fixed orientation of CGRO with respect to the sky (for a single
pointing period) generates continuously
varying backgrounds which range over a factor of two or more.  The lower 
energy background (up to $\sim$100 keV) is dominated by the modulation of
the diffuse sky flux by the Earth and at higher energies by the cosmic
ray secondary radiation as shown in Fig. 2.  Thus the low energy
background exhibits a slow sinusoidal variation 
with the orbital period of the spacecraft. At high energy, the background is
modulated
more rapidly due to the changes in magnetic field strength and direction.  
Furthermore, a given source exposes a combination of
several detectors at different angles.  The effective area of the
LAD (the product of the geometric area and the efficiency) is also 
a strong function of energy just above the lower energy threshold
due to the 
entrance window attenuation. It decreases more slowly at higher energies
from Compton leakage.  All these factors combine to produce complex
energy and time-dependent variations in the background.  Nevertheless,
we can use the known Crab Nebula flux as a standard candle to 
perform an empirical calculation of the sensitivity.

A basic representation of the
instrument sensitivity to a point source of continuum emission over
a specified energy range is
\begin{equation}
F_{min} = \frac{N_{\sigma}} {A \epsilon} \sqrt{\frac{R_{\rm B}} {T_{live}}}  
\label{eqn:min_F1}
\end{equation}
where $F_{\rm min}$ is the minimum detectable flux for
{\bf $N_\sigma$} standard 
deviations, {\it R$_{\rm B}$} 
is the background count rate,
$\epsilon$ is the 
detector efficiency, $A$ 
is the geometric area of the detector, and 
$T_{\rm live}$ is the livetime 
of the observation. 
This expression does not lend itself readily to an EOT sensitivity calculation, since
the background changes within the fitting window and 
$T_{live}$ is not a well-defined quantity.  In addition,
correction for a small rate-dependent electronic deadtime is made at that time 
for the CONT data in counts are converted to count rates.
Therefore we replace $\sqrt{R_{\rm B}/T_{live}}$ with the 
uncertainty $\delta r_{\rm s}$ from the least 
squares fitting problem (see Eq.~(\ref{eqn:rterr})), so that
\begin{equation}   
F_{\rm min} = \frac{N_{\sigma}}{A\epsilon} \delta r_{\rm s}.
\label{eqn:min_F2}
\end{equation}
Note that Eq.~(\ref{eqn:min_F2}) combined
with the semi-analytical expression Eq.~(\ref{eqn:rterr}) can be used to estimate Earth
occultation sensitivity for a given background rate; however, we can
use the Crab Nebula with known flux and measurement errors to obtain a
more accurate calculation.  

The Crab Nebula flux
($F_{\rm Crab}$) can be related to the measured Crab count rate 
$r_{\rm Crab}$ using 
\begin{equation}
r_{\rm Crab} = A \epsilon F_{\rm Crab} 
\label{eqn:rate_Crab}
\end{equation}
Combining Eqs.~(\ref{eqn:min_F2}) and (\ref{eqn:rate_Crab}) by eliminating
the area and efficiency factors
yields
\begin{equation}
F_{\rm min} = \delta r_{\rm Crab}
            \frac{N_\sigma F_{\rm Crab}} {r_{\rm Crab}} 
\label{eqn:min_F3}
\end{equation}
We adopt a best fit broken power law for the Crab Nebula spectrum
in
photons cm$^{-2}$ s$^{-1}$ keV$^{-1}$ obtained from HEAO-A4 measurements
(Jung 1989)
\begin{eqnarray}
\label{eqn:S_Crab}
{\cal S}_{\rm Crab}(E) = 3.25 \times 10^{-3}/(E/45)^{2.075} & 
\mbox{for $E < E_{\rm B}$} \\
\noindent\mbox{and} \nonumber
{~\cal S}_{\rm Crab}(E) = 3.732 \times  10^{-4}/(E/E_{\rm B})^{2.48} & 
\mbox{for $E \ge E_{\rm B}$} \nonumber
\end{eqnarray}
where $E_{\rm B}$ = 127.7 keV.
Equation~(\ref{eqn:S_Crab}) has been used previously in 
evaluating the performance
of BATSE prototype LADs in balloon observations of supernova SN 1987A 
(Pendleton \etal 1995b).  The HEAO A-4 spectrum compares well with the
observed Crab Nebula spectrum during the balloon flight.
We can now compute the sensitivity as a function of energy,
angle and detector combination, provided the source signal is adequated
determined. 

A large body of Crab Nebula occultation data was obtained in all eight
detectors in the 16 CONT data channels at many orientations of the CGRO,
covering a period of about five years.
These data, as we discuss later in the context of systematic error, 
were parameterized so that the Crab Nebula source rate at any
angle or energy in the LADs could be determined.  
The parametrization therefore automatically contains
the averaged exposure and background effects typical of BATSE data,
and can be used to determine $r_{\rm Crab}$ and $\delta r_{\rm Crab}$ in
eq.~(\ref{eqn:min_F3}).  

Using eq.~(\ref{eqn:min_F3}) and integrating eq.~(\ref{eqn:S_Crab}) over the
appropriate energy
range to obtain $F_{\rm Crab}$, we derive the point source sensitivity of a
typical two week observation (approximately
the length of one CGRO pointing period). This is shown as a function of energy and
angle in Figs. 7 and 8,
respectively.
As the angle to the source from the detector normal increases (Fig. 8), two,
three and four detector combinations in the octahedral geometry become
more sensitive than a single detector as shown.  
Table 1 compares the sensitivity of the flux extraction method and the
two point source search methods discussed later. 

It should be kept in mind that the sensitivity in
crowded sky regions such as the galactic center region is degraded from
that shown in Figs. 7 and 8 due
to additional terms in Eq. (\ref{eqn:mod_rate}) for known interfering sources
and systematic error in the flux measurements. 
We discuss causes of systematic error in the next section.

\subsection{Systematic Errors Related to Sky Location and Limb Geometry}

Several sources of systematic error were identified in the use of
the EOT to measure source fluxes.  We have
identified
those based on location in the sky combined with orbital precession effects,
unusual variations in the background, and also absolute flux using the 
instrument response.  The treatment of systematic error varies depending on 
use of the data. 

The current method of extracting flux and spectral occultation data assumes
some {\it a priori} knowledge of source intensity for sources other than
the SOI, i.e., a decision is made as to what
sources must be included in eq. (\ref{eqn:mod_rate}). The selection of which
sources are included in
the fitting window is therefore an important determiner of
the systematic error in two cases: (1) where residual sharing of flux occurs
between source terms in the fitting process, and (2) where sources are
neglected intentionally,
because of
an incorrect assumption about the sources' relative intensities to the
SOI,
or unintentionally, simply because an unknown source was present in the
fitting window at the
time.  It is important therefore to keep a controlled database of bright
source information
as discussed in Section 3.1.  Generally, sources that exceed $\sim$0.02
photons cm$^{-2}$s$^{-1}$
in the 20-100 keV band (about 75 mCrab for a source with a Crab-like
spectrum) are considered
as sufficiently bright to be included in the fitting window.  This
corresponds to about
2$\sigma$ detection in one day sampling by Earth occultation (see Table 1).
We have chosen a modest flux level that is comparable to our one-day
sensitivity.    Setting
thresholds in count space would be preferable, but this adds significantly
to the computational
burden.  A single threshold in photon space should be regarded as a first pass
method in an interative process designed to achieve best results. No account is made for the spectral differences between the
SOI and potentially interfering bright sources. It is possible,
for example, in a 
reanalysis to set finer intensity thresholds or make energy-dependent cuts.

The Crab Nebula and the
black hole system Cyg X-1 generally are the most persistently bright hard
X-ray
to low energy gamma ray sources in the sky.  Only occasionally do other
sources, usually transients, exceed the intensity of these objects.  The
Crab and Cyg X-1 are therefore dominant contributors to systematic error
in their respective sky regions when their occultation steps fall into the
fitting window
of the SOI.

An analysis of these errors can be made by examining the limb geometry as a
function of time.
In Fig. 9 we show the sky region near the
Crab Nebula and Cygnus X-1 and
surrounding sources monitored by BATSE. In each plot, two shaded regions are
shown.
Those bounded by dashed curves represent the set of projections of the
setting limb of
the Earth as the spacecraft moves through one precession cycle of its orbit. 
The shaded regions bounded
by solid curves are the equivalent regions for the rising limb.   The limb
appears noncircular because of the flat sky projection. 

Note that some regions of the sky are not swept by
the Earth's limb at times of occultation of the Crab and Cyg X-1.  
For example, the gamma ray sources GRO J0422+32 and 
Geminga are located in sky regions that are not crossed by the projected
limb of the Earth
at times of occultation of the Crab.  Light curves generated for either
these two sources show
minimal deviation with precession phase due to the presence of Crab
occultation steps near in time to those of the two sources of interest.  In
contrast, 4U 0614+091
or SAX J0501+11, are located such that the projected setting and rising
limbs of the Crab, 
respectively, cross them at certain times during the precession cycle.  This
is equivalent to the Crab
and the two sources having occultations at essentially the same time for a
period of 1-2 days.  Light
curves for these two sources will     
exhibit significant systematic deviations in their light curves near and at
these times.   

This is illustrated in Fig. 10, where we show the time dependence of the
angle between the locations
of the sources (a) GRO J0422+32 and (b) 4U 0614+091, and
the closest approach points of the Earth's limb 
projection that intersect the Crab Nebula over a period of 200 days, or
about four precession
cycles. The corresponding light curves over the same time intervals are
shown.  GRO J0422+32
is not detected during this time, but 4U 0614+091 is detected consistently,
but
variable ($\sim$0.01 photons cm$^{-2}$ s$^{-1}$).
The minimum angle for the setting limb of
Crab  approaching GRO J0422+32 is about 3$^{\circ}$, 
and therefore the occultation steps are well separated.  However, for
the setting limb of the Crab crosses the location of
4U 0614+091, and the occultations become superposed for a few days every
precession cycle
(for example, around TJD 9900).  The corresponding
light curve for 4U 0614+091 shows systematic deviations at these times due
to 
residual sharing of source signal as we approach the conjunction of the 
setting occultation steps.  Our general treatment of these light
curves is to reject the measurement of occultation steps 
of 4U 0614+091 that
fall within 10 secs of the Crab steps ($\sim$0.7$^{\circ}$).  The amount of
signal
sharing depends on factors such as the broadness of the occultation
features (governed by the elevation angle, $\beta$, of the source above
the CGRO orbital plane, see Appendix B) and the steepness of the
background.  Sometimes a cut is made on the
step measurements when the angle between the limb and the source location is
less than 2$^{\circ}$, which unfortunately creates or increases gaps in
coverage when
both the rising and setting limbs have poor geometries for separating the
flux
from the SOI from interfering sources.
Here we derive an average systematic error as discussed below to account
for this effect.

Our investigation of several sources of systematic error illustrates that it
is difficult to quantify the total systematic error precisely.  Unknown 
systematic
errors due to source confusion are time and sky location dependent.  An
additional
source of systematic error, discussed in the context of source step
searches,
is the presence of non-statistical background fluctuations (red noise
and coherent pulses) from bright X-ray sources.
We generally characterize these
as backgrounds components that are, in practice, unpredictable.

Another source of error that we have not treated rigorously is
the diffuse galactic emission along the galactic plane
(Valinia \& Marshall 1999). 
Under appropriate limb conditions, the diffuse component
may have a sufficiently sharp profile
in galactic latitude at energies in the sensitive range
of BATSE to create occultation
step-like features in the background and/or to increase the amount of
cross-coupling between
background and source terms in the fit.  At this time, we have not tried to
separate 
the diffuse component from possible weak point source contamination in the
galactic bulge region.
 Although the presence and structure of the ridge
is scientifically interesting, modeling of this component is best done in
context of a global background model, and so beyond the scope of this paper.

We wish, nevertheless, to estimate a total {\it average}
systematic error (source confusion + unpredictable
background variations).  To do this, we created a 
grid covering the galactic plane 
in areas where sky-dependent systematic error is a significant 
problem as shown in Fig. 11.   The grid has points every 3$^{\circ}$ along
the galactic
plane with two sets of grid points at +6 and -6$^{\circ}$ galactic latitude.
Grid points 
within 2$^{\circ}$ of known bright
occultation sources were excluded, yielding a total sample of 156 points.
 
In Fig. 12, we show the average flux from each grid point averaged over
a 7.2 year period and the standard deviation of one-day flux averages from
the beginning of the 
CGRO mission, in 1991 April, to 1998 July, as a function of galactic
longitude, ${\it l_{II}}$. 
The standard deviations are clearly broader than predicted
for a Gaussian distribution about zero flux due to systematic errors
from various effects (1 $\sigma$ = $\sim$0.01 photons cm$^{-2}$s$^{-1}$).  
We find broadening factors of about 30\% to 60\% in excess of normal
statistics, strongest near the galactic center,
and near bright, variable sources such as Vela X-1
($l_{II}$=-96.9$^{\circ}$) and
Cygnus X-1 ($l_{II}$=71.3$^{\circ}$). There are also small positive and
negative
trends in the average fluxes for the 156 test points. Absolute 
values are less than $\sim$0.01 photons cm$^{-2}$s$^{-1}$, with the
exception of the galactic center itself.  Even though small,
the galactic bulge region
between
+60$^{\circ}$ and -60$^{\circ}$ shows a clear effect, presumably 
due to weak X-ray and gamma ray sources in greater numbers, and/or a
galactic
ridge component. The trends for galactic latitude are similar, but narrower
in spatial extent.
These results
are used to estimate systematic error in the 
BATSE Earth occultation catalog of low energy gamma ray sources (Harmon
\etal in preparation).

\subsection{Systematic Errors Due to Detector Response Model}

The response matrices for the BATSE large area detectors have been
described
elsewhere (Pendleton  \etal 1995a), as well as their use for locating gamma
ray bursts
(Pendleton \etal 1999).  The matrices are an ordered, discrete representation of the
BATSE gamma-ray detectors' response characteristics, and thus are used for
all types of studies such as bursts, distant point sources, solar flares and
atmospheric gamma ray phenomena.  They are designed to convert
background-subtracted source counts to incident photon spectra.  The main
parameters used to characterize the response is the incident photon energy, the
measured detector output energy, and the angle between the detector normal and
the source direction.  LAD prototypes were used in balloon flight observations
of SN 1987A and the Crab (Pendleton \etal 1995b), for which an early version of
the matrices eventually used for BATSE was developed.  A Monte Carlo simulation
was used to generate the BATSE LAD responses based on a well-known
electromagnetic cascade/transport code called EGS (Ford \etal 1978; 1985) and a
mass model for the LAD, its mount, and more crudely, for the remainder of the
CGRO spacecraft.  Prior to launch, the LAD modules were calibrated using angular
response and absolute efficiency test data taken with radioactive sources 
(Lestrade 1989, 1991; Horack 1991).  The Monte Carlo simulations were then optimized using
the test data.

Scattering of source radiation off the upper atmosphere can produce substantial
fluxes in near Earth-pointed detectors, when a point source is well above the
horizon.  For example, scattered radiation from an overhead gamma ray burst can
contribute to the measured flux by as much as 50\% (Pendleton \etal 1999). 
However, for Earth occultation, which is near line-of-sight to the horizon for
the source direction, our simulations indicate that the atmospheric scattering
is negligible.  Therefore only the direct response matrices need be used for
deconvolving fluxes as described here.

In the course
of our science analysis of Earth occultation data, two significant
systematic
effects were found that could be traced to the response model: (1) the CONT
channel boundaries
were difficult to calibrate using a simple function of energy, especially
below 100 keV,
due to nonlinearities in the onboard analog
electronics.  This created unacceptably large
residuals in spectral model fits and model-dependent flux measurements, and
(2) the instrument response for near
face-on
flux measurements was under-predicted in the model.  This caused fluxes for
sources near the normal direction to a LAD, less than about 20$^{\circ}$, to
be 
artificially high by as much as 30\%.  Unfortunately, these two sources of
systematic error were not understood prior to the launch of CGRO, and
corrections to the response using in-flight data were complicated due to
the coupling of the two effects.   

Preece \etal (1994) and Pendleton \etal
(1994) have
used various data types to better determine the effective energies of the
CONT channel
boundaries.  Each approach is different, and, although 
neither appears to be superior to the other, both schemes 
are an improvement over the pre-flight calibration (Lestrade 1991),
which makes no adjustments for the electronic nonlinearity. Both the Pendleton and
Preece 
algorithms are incorporated into the analysis of BATSE occultation data.
For this work,
all spectra and light curves were determined using the Pendleton algorithm
which optimizes
the channel boundaries using Earth occultation and pulsar data in a joint
analysis
(Pendleton \etal 1994).

To more fully investigate the LAD response as a function of photon energy
and source aspect angle, i.e., the angle between the detector normal and 
the source direction,
the Crab Nebula occultation dataset for all eight
detectors in the 16 CONT data channels at many orientations of
the CGRO was used.
An example of these data is shown in Fig. 13 for 
CONT channel 3 (40-50 keV) as a function of angle in LADs 0, 1, 2, and 3.
We find that the Crab count rate is a smooth function of angle and highly
reproducible.
This is consistent with the Crab Nebula flux being
constant to within about 5\% fractional rms
for the 20-100 keV band over the 5-year sampling period (see Fig. 14). 
The dotted curve represents the
modeled response of the known Crab Nebula flux (eq.~(\ref{eqn:S_Crab})),
before (four top frames) and after (four bottom frames), response
corrections (described below) have 
been applied,
and the solid curve (same in top and bottom frames) represents an empirical
fit used to characterize
the Crab rates as a function of energy and angle 
(Laird \etal 1996).  For all energy bands, and at
intermediate aspect
angles ($\sim$20$^{\circ}$-70$^{\circ}$), the agreement between data and 
the response model is excellent.
However, in the lowest CONT channels (1-3), as in part
(a) of Fig. 13,
the model underpredicts the actual measured Crab rate as we approach the
normal direction to the detector face, yielding
a flat response at angles of 20$^{\circ}$ or less, whereas the true
response is more forward pointed as indicated by the data.  
This effect was partially attributed to the simplifying assumptions made in
the LAD
response model for the detector entrance window.  

The entrance window consists of two 0.635 cm layers of a light weight 
aluminum-epoxy composite, called HEXEL, interleaved with other low Z
(charge) materials to minimize
attenuation of the low energy photons.  
 The HEXEL provides support and optical insulation for a plastic
scintillator in front of the NaI
crystal for charged particle rejection.  Viewed from the direction of
incoming photons, the HEXEL
appears as a close-packed hexagonal array, or ``honeycomb" of cells.  The
cells are 
about 0.3-0.5 cm in diameter, and irregularly shaped with no alignment of
cell walls between
the two layers, creating a difficulty for 
Monte Carlo modeling. In the original response matrix generation (Pendleton
\etal 1995a),
the HEXEL is treated as uniform layers of aluminum of reduced effective
thickness.

During the mission, we surmised that the entrance window
allowed more flux into the detector than was predicted in the Monte
Carlo model of the LADs. The aluminum-epoxy is very thin in the direction of
the normal to the detector face, but thickens faster than 1/cos$\theta$ as
the angle increases, acting as a low energy collimator. We therefore
performed calibration tests of an entrance window for the LADs
at Eastern Kentucky University (EKU) (Laird \etal 1996) to 
quantify this effect.  The results revealed
an energy and angular dependence due to the double HEXEL layer consistent
with the known composition of the material. 

The HEXEL was found to account, however, for only a portion of the observed
forward-angle response deficiency and only in the lowest three energy bands
(20-60 keV).
An additional detector and
angular dependent effect was observed between the Crab data and
and model predictions over the angular range 
(0$^{\circ}$-20$^{\circ}$) and the entire 
energy range (20-2000 keV) of the LADs.  We attribute the additional
deficiency
to the adoption of a power series in cosine of the aspect angle (Pendleton 
\etal 1995) that was
used to interpolate response matrices between the Monte Carlo results at a
predetermined set of angles.  This function (dotted curves in part (a)) is
flatter in the
forward direction than the Crab data predict as shown in Fig. 13.  A more
physical function,
including the triangular response of the HEXEL at low
energies and the cosine of the geometric area (Laird \etal 1996), was used to fit the data
(solid curves in part (a) and (b)) instead.
  
Ideally, it would be best to regenerate the response matrix model with 
the forward angle response properly treated, but operationally it is 
more convenient to consider the effects discussed above as additive,
and determine an effective correction to account for the discrepancy
of model and data.

Using the empirical fits as shown
in Fig. 13 as a benchmark, we derived total corrections to the response
model, including both the entrance window and empirical angle corrections.
The response matrix is multiplied by a factor of
the form
\begin{equation}
A \exp(-B\theta-2\mu\tau_{eff}+2\mu\frac{t}{cos\theta})
\label{eqn:corr_fac}
\end{equation}
prior to spectral fitting.
Here $\tau_{eff}$ represents the effective thickness of the HEXEL derived
from the EKU measurements,
and $\mu$ is the energy-dependent attenuation coefficient for aluminum.
The constant A is determined by matching the model and data for each 
detector at 0$^{\circ}$. The constant B, which determines the 
sharpness of the 
angular correction, was adjusted to reduce the residual difference between
the dotted and solid curves
in part (a) of Fig. 13.   Neither A or
B are energy dependent, and B has the same value for all detectors. The
corrections given in eq.~(\ref{eqn:corr_fac})
have been incorporated into
all spectra and intensity histories presented in this paper and in 
the catalog of low energy gamma ray sources (Harmon \etal in preparation).  

A histogram of the average daily Crab Nebula fluxes
as a function of time before and after the correction is shown in Fig. 14.
The histogram of flux measurements after correction is more symmetric and
less-skewed toward higher fluxes.
The fractional rms of the measured Crab flux has dropped from about 5.2\%
to 4.3\% in the 20-100 keV band.   

\subsection{Absolute Flux Calibration}

We attempt here to perform an absolute calibration of the
Earth occultation flux measurements using the Crab Nebula emission as
a standard candle. We break the absolute flux determination
into two parts.  First, we compare Earth occultation measurements
of the Crab Nebula
with lunar occultations of the same source to avoid any systematic effects
due to using the
Earth and surrounding atmosphere as an occulting disk, and also independent
of the BATSE 
response model. Second, we compare observations
of the Crab Nebula with other high energy instruments, assuming that
the Crab is a steady source of high energy emission.  

From 1993 April to 1994 January, the Crab Nebula was occulted several times
by 
the Moon. Since the Moon has no atmosphere, lunar occultations can in
principle
be used to measure the absolute Crab intensity without the systematic
uncertainty
due to atmospheric scattering. 
To find when the Crab was being occulted by the Moon, we first searched for
minima
in the angle between the Earth-Moon vector and the unit vector to the
Crab. Next we did more detailed calculations to see if the Crab was being
occulted by the Moon on these days. If it was not Earth occulted, we
computed the inner product
$\lambda(t)$
\begin{equation}
\lambda(t) = \mathbf{{\Delta x}} \cdot \mathbf{{\Omega}} = 
|\mathbf{{\Delta x}}| \cos \alpha
\end{equation}

where  $\mathbf{\Delta x}$ is the vector from the {\em CGRO}
spacecraft to 
the
Moon, calculated every 2.048 seconds using the spacecraft position in the
BATSE
housekeeping data and the Moon position from the complete JPL DE-200 
ephemeris (Standish et al. 1992).
$\mathbf{\Omega}$ is a unit vector pointed from the Moon in the
direction 
of the 
Crab Nebula, and $\alpha$ is the angle between these two vectors, as shown
in
Fig.~\ref{fig:crab_moon_geometry.ps}. If
$\lambda(t) \leq 0$ then a lunar occultation is possible. The Crab is
occulted
by the Moon if
\begin{equation}
\rho^2 = |\mathbf{{\Delta x}} |^2 - \lambda^2 < R_{\rm Moon}^2
\end{equation}
where $R_{\rm Moon} = 1737.4$ km is the radius of the Moon. The time of
occultation is computed by linear interpolation between BATSE housekeeping
records. The times of lunar occultations of the Crab and the BATSE LADs
(0-7) that 
observed them are listed in Table~\ref{tbl:moon_occ_times}. 

To find the intensity of the Crab Nebula, lunar occultations were fitted in
each
energy channel and each detector with a model of same form as
Eq. (\ref{eqn:mod_rate}), except that we replaced the atmospheric
transmission with a step model for the Crab-Moon occcultation as follows.
We assumed the Crab Nebula to be an extended source with uniform emission,
i.e.,
\begin{equation}
T_n(t) = \left\{ \begin{array}{ll}1 & \mbox{if $t<t_{\rm occ}-\tau/2$} \\
                                  \pm\frac{1}{\tau}(t-t_{\rm occ})
				  +\frac{1}{2} & 
				  \mbox{if $t_{\rm occ}-\tau/2 \leq t \leq
				            t_{\rm occ}+\tau/2$} \\
                                  0 & \mbox{if $t>t_{\rm occ}+\tau/2$} 
                 \end{array} \right.
		 \label{eqn:moon_occ_ext}       
\end{equation} 
Positive and negative signs indicate setting and rising
occultations, respectively. $t_{\rm occ}$ is the occultation time, and
the width of the occultation step in time  
$\tau$ is estimated as 
\begin{equation}
\tau = \frac{D_{\rm Crab}}{D_{\rm Moon}} \Delta t
\end{equation}
where $D_{\rm Crab}$ is the angular extent of the Crab Nebula, 
$D_{\rm Moon}$ is the maximum angular extent of the Moon (1900\arcsec), and 
$\Delta t$ is the total time the Crab is occulted by the Moon. 
For each pair of steps (a set and a rise), both steps were fitted
simultaneously.
Background data were used from 110 seconds before the first step fitted
until 
110 seconds after the last step fitted. An example of a fit with this model
is 
shown in Fig.~\ref{fig:channel3_lad7_extend.ps}. 

Using a grid search to find the best value of $D_{\rm Crab}$, we minimized
\begin{equation}
\chi^2 =  \sum_{p=1}^M \sum_{k=1}^{N^p} \sum_{j=1}^{N_{\rm det}^p}
 \sum_{i=1}^{N_{\rm chan}} 
 \frac{(r_{ijkp}-y_{ijkp}(t_{kp}))^2}{\sigma_{r_{ijkp}}^2}
\end{equation}
where $M$ is the number of pairs of rises and sets; $N^p$ is the number of
data
points for pair $p$; $N_{\rm det}^p$ is the number of detectors viewing pair
$p$; $N_{\rm chan} = 10$ is the number of energy channels used; $r_{ijk}$ is
the BATSE count rate in energy channel $i$ and detector $j$ at 
time $t_k$; and $y_{ijkp}(t_{kp})$ is the count rate model
evaluated
at time $t_{kp}$ for energy channel $i$, detector $j$, and pair $p$.   
The energy channels used correspond to 20-300 keV. Our best fit to the
angular extent of the emission region of the Crab Nebula, assuming uniform 
emission, was $D_{\rm Crab} = 131\arcsec\ \pm 5\arcsec\ $. This agrees well
with
previous results from lunar occultations of the Crab (Staubert \etal 1975). 

To compare BATSE Earth and Moon occultation measurements of the Crab, we
used 
the characterization of the BATSE Earth occultation count rates for the Crab
as
a function of energy, aspect angle, and detector from empirical fits as
shown in Fig. 13
with our fits to lunar occultations. 
Figure~\ref{fig:moon_earth_compare_extended.ps} shows the comparison between
lunar occultations
with our extended source model (Eq.~\ref{eqn:moon_occ_ext}) and Earth
occultation measurements. We find that the Crab flux values from Moon 
occultation are on average about 95.5($\pm$1.9)\% of those measured with
Earth 
occultation. Although the results we have obtained depends on the assumed
model 
of the Crab Nebula emission, we believe that the error due to this effect is
small. The Earth occultation fluxes are systematically higher by a few
percent.
This could be due to the simple atmospheric model we use, which neglects 
scattering of source flux off the atmosphere away from the line-of-sight.
Atmospheric scattering is substantial at more oblique angles to the
atmosphere,
and is a significant effect in determining GRB locations (Pendleton \etal
1999).
For near line-of-sight scattering into the FOV of the detector at rise or
set, 
however, we do not calculate significant contributions to account for the
observed effect. 

Observations of the Crab Nebula spectrum using a simple broken power law
model for
fitting data between 20 keV and 1 MeV 
have been performed with a Naval Research Laboratory balloon experiment
(Strickman, Johnson, \& Kurfess 1979), HEAO-1 (Jung 1989), the Oriented
Scintillation
Spectrometer Experiment (OSSE) on board the CGRO (Much \etal 1996;
Strickman,
private communication) and the Gamma-Ray Imaging Spectrometer (GRIS) 
(Bartlett 1994).  In Fig. 18, we show the measured 
BATSE spectrum of the Crab Nebula compared to these measurements
based on the broken power law fit, and a confidence region obtained by
varying the
high and low energy spectral indices, and the break energy separately by
three $\sigma$
(see Eq.~(\ref{eqn:S_Crab})).  The spectrum is a composite result 
of fitting each of the eight LADs simultaneously at times when the Crab
aspect angles
are 30$^{\circ}$ or less.  The BATSE response has been corrected according
to the prescription
given in Eq.~(\ref{eqn:corr_fac}) and reduced by 4.5\%, consistent with the
absolute calibration by Moon
occultation. We also show RXTE Proportional Counter Array (PCA)
data in the 10-30 keV band, RXTE High Energy X-Ray Telescope (HEXTE) data in
the 20-200 keV band (Heindl, private communication) and BeppoSax Wide Field
Camera (WFC) data (In't Zand, private communication) in the 10-25 keV band.
The BATSE data agree reasonably well with most of the other 
measurements of the Crab Nebula spectrum below $\sim$200 keV, except OSSE
and GRIS.  
 
The discrepancy between BATSE and OSSE is quite significant
below $\sim$150 keV (15\%-20\%) and has been documented previously by Much 
\etal (1996).  No adequate explanation for this has resulted from our analysis. 
We have eliminated the possibility of atmospheric scattering, and corrected
systematic errors due to the response model.  In fact, a normalization of our
data to the OSSE results at 100 keV (difference is about $\sim$16\%)
forces the effective area of the LAD ($\sim$1700 cm$^{2}$)to
be roughly the same as its geometric area (2025 cm$^{2}$)!  A comparison to other published instrumental 
results, with the exception of GRIS (Fig. 18), suggests that the OSSE response 
at lower energies (less than
about 150 keV) is overestimated.  The OSSE Crab Nebula spectrum tends to
increase the change in photon spectral index from above and below the break
energy (120-200 keV).  The GRIS balloon-borne germanium detector also shows a breaking spectrum in this energy
range, but the discrepancy of these data with BATSE is the largest of 
all instruments.  Again, the reason is unknown, but underscores the difficulty
in precision normalization of gamma ray astronomy telescopes which typically
involve complex background and detector response effects.  In our
BATSE calibration exercise, we have looked for overall consistency between
the eight LADs and compared the basic mechanism of occultations by the atmosphere
to those by the Moon.

\section{SEARCH TECHNIQUES FOR UNKNOWN SOURCES}

\noindent
\subsection{Single Step Searches}

After the data selection process as described in Sec. 2.2, a 
daily search of BATSE CONT data was routinely performed to find 
occultation features of the brightest hard X-ray/gamma
ray sources below the gamma ray burst trigger
level.  The sensitivity of the search method depends on a variety of factors
such as the shape of the source spectrum, energy bandpass,
the number and orientation of detectors with respect to the source used,
and the behavior of the background.   Sources that reached levels
of about 4.6$\times$10$^{-9}$ erg cm$^{-2}$ s$^{-1}$ (about 200 mCrab for
sources
with Crab-like spectra)
are detectable with the step search method (see also Table 1).  

Data in each LAD are first summed over the CONT
data channels where the best signal-to-noise (S/N) ratio for sources with
Crab-like spectra is obtained (30-200 keV).
A series of fitting tests are then performed on a contiguous data
segment within a fitting window of four minutes
centered on a test time {\it t$_{occ}$}. We then determine if an 
occultation step is present in the background somewhere
within the window.
Once a segment has been tested, positive or negative
results are logged for post analysis, and the window
moves or ``slides" to the next data segment.
To maximize the sensitivity
of the sliding search, a grid of points along the Earth's limb
is calculated at {\it t$_{occ}$}.  Using a cosine function to represent 
the response of the detector, the combinations of the eight LADS 
(out of a possible 28) with the best S/N ratio for each
point are determined.  These combinations are then tested at the
corresponding limb points.

The first test is
a linear fit over the fitting window for each LAD combination
to locate deviations from a background
of approximately constant slope.  If the goodness of fit ($\chi^{2}$) 
per degree of freedom is above a threshold (usually set to $\ge$1.0),
a second fit is performed over the same fitting window using
a template step function to determine if the deviation is
caused by an source occultation ($\chi^{2}$
required for an occultation step is typically $\le$1.3).  
A third fit is then made to estimate the best time {\it t$_{occ}$}
of the occultation and the maximum significance of a step.  The significance
cutoff is set to a minimum of 2 $\sigma$ to avoid large numbers of
spurious step detections. 

The results of an occultation search for a single day are shown in Fig. 19.   
The significance of detected occultation steps is plotted as a LAD
number (0-7) as a function of time elapsed from the beginning of the
spacecraft orbit for each orbit searched. The plotted symbol corresponds to
the LAD number (0-7) with the most significant step in the combination of
detectors.
The time zero is the
time of the first data packet for the day.  Plotted in this way, 
detected steps from the same source cluster at a specific time, although a
shift of a few seconds 
occurs in
the occultation time folded on the orbital period due to the orbital
precession, which is about 0.5$^{\circ}$ per orbit.  The times of
bright source occultations, such as those for the Crab and Cygnus X-1,
are shown on the plot.  Significances above
zero  and below zero indicate rising steps and setting
steps, respectively.  

This method allows a quick determination of occultation times
of bright transient or flaring sources from which the location of a source
can be deduced from the orientation of the Earth's limb at rise and
set.  In Appendices B and C we describe a graphical method for localizing a 
point source based on the time of occultation steps. 

The single step search/graphical method for determining source locations was 
used primarily in the early part of CGRO mission, prior to 1993, after which we
developed Earth occultation imaging to improve our ability for source
localization.  Two effects severely limited the sensitivity of the single
step search.
In the first place, orbit to orbit variations in the background make
it impractical to combine raw data for different orbits. 
A change in the
background of a few percent from one orbit to the next
can induce
features in the background that mimic occultation steps.
Without a reliable
predictor of the background or a de-trending algorithm, orbit by orbit
data cannot be folded to increase the sensitivity (this is overcome
in the imaging technique discussed in the next section).
The second problem is the presence of sub-threshold
transient features that escape the manual flagging procedure or are
impractical
to flag.  An example is the presence of bright long-period
(P$\sim$minutes) pulsar signals
in the background data.  Pulses from sources 
such as Vela X-1, GX 301-2 or other bright high mass binary
transient systems are a source of nonstatistical noise. 
This problem was partially overcome by
increasing the length of the fitting window and the bandpass for the search,
at the expense of sensitivity
to softer spectrum sources. 

In general, the accuracy of the EOT for determining a source's sky
position
depends very strongly on the orientation of the Earth's limb at the time of
detection.  If the rising and setting limbs are nearly parallel, which is true
when the source direction is very close to the orbital plane of the spacecraft,
it is obvious the positional accuracy along the limb direction can be very poor.
On the other hand, allowing orbital
precession to move the limb geometry into a more favorable configuration,
can greatly improve positional accuracy.  In a situation of optimal
limb geometry, the width of the occultation feature itself and the
S/N ratio limits the ultimate positional accuracy.
The spatial resolution perpendicular to the limb is
determined by the width of the occultation step in time, which takes about
8/cos $\beta$ seconds, where $\beta$ is the angle of the source with respect
to the orbital plane (see Appendix B). For a 90 minute orbit, the angular
resolution on the sky is (8s/5400s)*360$^{\circ}$, or $\sim$ 0.5$^{\circ}$.
Table I gives a lower limit
on the positional error, considering these factors.  For the
step search method, the best obtainable accuracy was $\sim$1$^{\circ}$
with a simple step function consisting of a fitted line for the background
and a steeper line with a step width of 10 seconds.
We were able to improve our locational accuracy to $\sim$0.2$^{\circ}$
by comparing
the step search results to the predicted shape of the Earth occultation
features from the atmospheric 
attenuation model (Sec. 3.1).  Usually a few iterations on the source
location were required to achieve the
best agreement with the data that were available at that time. Examples of the
use of this method 
are given in Harmon \etal (1992), Paciesas \etal (1992) and Harmon
\etal (1993), for the
sources 4U 1543-47 (0.25$^{\circ}$), GRO J0422+32 (0.68$^{\circ}$), 
and GRO J1719-24 (=GRS 1716-249) (0.14$^{\circ}$), respectively, where
the number in parenthesis is the difference in degrees between
the reported BATSE location and that of the optical counterpart from the
electronic database SIMBAD.        

\noindent
\subsection{Occultation Imaging}

Occultation features in the time domain can be transformed into spatial
information for construction of images.  This enhanced the ability
of BATSE to locate and identify much weaker sources than the step search
algorithm permitted, and considerably increased the effectiveness of the
instrument
as an all-sky monitor (Zhang \etal 1993, 1994a).  It also allowed us to obtain
positional information more efficiently and easily for new transient sources and 
to 
discriminate between sources in crowded regions such as the galactic center.  

Occultation transform imaging is conceptually similar to techniques 
used for many years in
radio astronomy (Bracewell 1956)
and medical X-ray tomography (Gullberg \& Tsui 1989) to convert
essentially one-dimensional scanning
measurements into two-dimensional images.
Using the Earth as a stable occulting disk,
the limb is projected onto the sky.  As the limb of the Earth (an arc,
not a straight line) sweeps through a chosen region of the sky, the LADs
record the counting rate as a function of time.  
The location of the spacecraft is
well-known and thus the location and orientation of the limb projected on
the sky can be determined accurately.  For point sources along the limb, a
change occurs in the detector counting rate as shown
in Figs. 2 and 3. For a single location on the 
sky, the intersecting Earth's limb
will change its angle due to precession of
the spacecraft orbit.  If enough one-dimensional strips
are sampled, an image can be generated.

Image reconstruction is achieved via the Radon transformation 
(Deans 1983) and the Maximum Entropy Method (Huesman \etal 1977).
Use of the Radon transformation to represent the limb projection limits 
sky images to
a field-of-view (FOV) of about 20$^{\circ}$ by 20$^{\circ}$.
Larger images produce distortion of point source 
locations near the edges of the FOV.  

We begin with a forward transform from
the image space (sky pixels) to the data space (count rate vs. time), 
which is accomplished in two steps.  
First, a curved Radon
transform is applied (Deans 1983) for the defined FOV.  This
is performed 
numerically since
no analytic form is available to describe the limb arc on the sky.  In the 
second
step, a high pass Butterworth filter from the Radon space to the data space
is executed
using a Fast Fourier Transform (FFT).  The Butterworth
filter was chosen over a differential filter (Zhang \etal 1995) used
in a first generation of the imaging system (Zhang \etal 1993) 
since it gives a better S/N ratio, 80\% vs. 50\%, and was also
less sensitive to spikes in the background data. 
The data are then filtered
with the same algorithm, however, the data (DISCLA)
usually contain
gaps due to flagging of fast timescale variations in the background, 
SAA passages
or loss of telemetry.  The best method we have found to fill data
gaps is to make a simple linear interpolation.  The result is a flattened
data residual with bi-polar shaped features of finite width 
for point source occultations with the slowly-varying background components
removed (Paciesas \etal 1995).  

Finally, the data space to image space reconstruction
is performed using the Maximum Entropy Method (MEM) (Gull \& Skilling 1984).
The backward transform
is physically a smearing of the original image with the forward transform 
function, and thus defines the direction in which the iterative process
converges.  The Maximum Entropy Principle is used as a stopping criteria
since an exact reconstruction is not achievable due to the presence of noise
and imperfect sampling forced by the available limb projections. 

Monte Carlo simulations have
been performed to confirm the reliability and sensitivity of the images 
(Zhang \etal 1995).  A simulated image and occultation data (after
filtering)
for two closely-spaced point sources using data which contain rises and sets
separated by two days is shown in Fig. 20.  

Examples of images are shown in Fig. 21 for sources along the galactic
plane.
The location accuracy near the center of a 
10$^{\circ}$ $\times$ 10$^{\circ}$
image is $\geq$0.3$^{\circ}$ for a best case geometry 
(nearly perpendicular limb samplings), with 
a sensitivity in the 20-100 keV range
of about $\sim$1.1$\times$10$^{-9}$ erg cm$^{-2}$ s$^{-1}$ (75 mCrab for a
source with
a Crab-like spectrum) for a one day integration.  Computer memory
requirements
limit the number of days that can be combined into one image, although
we have achieved integrations of 15-20 days, with a sensitivity between
10 and 20 mCrab.  For locating a new source,  generating an image
such that a source is near the center of the FOV, produces a sensitivity and 
positional accuracy that is close to
the standard flux extraction method (see Table 1).  The chief
advantage of imaging method is that it requires less {\it a priori} knowledge
of a new source location and the details of
orbital precession and the presence of interfering sources are naturally taken
into account.   Thus the method allowed deep imaging of short-lived 
transients (a few days), and identification
of weak high energy emission from low mass X-ray binaries (Barret \etal
1996), active 
galaxies (Malizia \etal 2000) and supernova remnants (McCollough \etal
1997).  Examples of new transients located with this method are GRS 1009-45
(Zhang \etal 1993) (0.39$^{\circ}$) and GRO J1655-40 (Zhang \etal 1994b)
(0.31$^{\circ}$), where the number in the parenthesis is difference in
degrees from the BATSE-reported location of the source and that of the
optical counterpart from SIMBAD.

\section{CONCLUSION}

We have shown various aspects of the Earth occultation technique
for point source studies with the BATSE large area detectors.  The technique
is applicable to all-sky monitoring or obtaining time- and
energy-dependent information about hard X-ray and low energy gamma ray
sources, and yields positional information for point sources
with moderate accuracy.
It has been used successfully for several years for independent science 
investigations and
triggering observations with other satellites and ground-based instruments.
The method can be used with an uncollimated detector and
without sophisticated background models.  Improved background modeling,
however, would allow one to increase the amount of data within the fitting
window and
abandon the simple polynomial fits, thus potentially increasing
the sensitivity of the method.  Several efforts have resulted in models
that can predict backgrounds to the few percent level (Ling \etal 1996;
Rubin 
\etal 1996b), but in practice, extracting point source fluxes in
the few mCrab range using BATSE requires sub-percent level accuracy.  

\section{ACKNOWLEDGEMENTS}

The authors would like to thank the BATSE Mission Operations Team for their
help in preparing software
and other data products, and especially Maitrayee Sahi, Burl Peterson and 
William Henze. B.A.H. is also especially grateful to William A. Wheaton, 
James C. Ling, and Duane Gruber for sharing their vast knowledge of
HEAO data analysis and studies of the high energy background in gamma ray
detectors.  We thank Chris Laird and students at Eastern Kentucky University
for measurement and characterization of the LAD entrance window.  
We also thank the following individuals for
instrument data on the Crab Nebula:
Mark Strickman, for OSSE, Keith Jahoda, for RXTE/PCA, Jean in't Zand, for
BeppoSax/WFC,
and William Heindl, for RXTE/HEXTE.  This made research has made use of the
SIMBAD database, operated at CDS, Strasbourg, France. This work was supported by the Goddard
Space Flight Center Compton Gamma Ray Observatory Science Support Center, which
is funded by the NASA Office of Space Science.

\newpage

\begin{appendix}
\section{LIST OF ACRONYMS AND ABBREVIATIONS}
\begin{table}[!ht]
\tablenum{A1}
\tablewidth{0pt}
\tablehead{\colhead{Acronym or Abbrev.} & \colhead{Meaning}}
\begin{tabular}{ll}
BATSE & Burst and Transient Source Experiment \nl
CGRO & {\it Compton Gamma Ray Observatory} \nl
CONT &  large area detector continuous data \nl
DISCLA & large area detector discriminator data \nl
EBOP & Enhanced BATSE Occultation Package \nl
EGS & Stanford electromagnetic cascade and transport code \nl
EKU & Eastern Kentucky University \nl
EOT & Earth Occultation Technique \nl
FOV & field of view \nl
GEANT & Southampton electromagnetic cascade and transport code \nl
GRB & gamma ray burst \nl
GRIS & Gamma-Ray Imaging Spectrometer \nl
HEAO & {\it High Energy Astronomy Observatory} \nl
HEXEL & honeycomb-like aluminum-epoxy composite \nl
HEXTE & High Energy X-Ray Timing Experiment \nl
ICAO & International Civil Aviation Organization \nl
INTEGRAL & {\it International Gamma-Ray Astrophysics Laboratory} \nl
JPL & Jet Propulsion Laboratory \nl
LAD & large area detector \nl
LEO & low Earth orbit \nl
MEM & maximum entropy method \nl
MJD & Modified Julian Date (Julian Date - 2,400,000.5) \nl
MSFC & Marshall Space Flight Center \nl
\end{tabular}
\end{table}
\begin{table}[!ht]
\tablenum{A1}
\tablecaption{List of Acronyms and Abbrev. (cont)}
\tablewidth{0pt}
\begin{tabular}{ll}
\tablehead{\colhead{Acronym or Abbrev.} & \colhead{Meaning}}
OSSE & Oriented Scintillation Spectrometer Experiment \nl
PCA & Proportional Chamber Array \nl
RXTE & {\it Rossi X-Ray Timing Explorer} \nl
SAA & South Atlantic Anomaly \nl
SD & spectroscopy detector \nl
S/N & signal-to-noise \nl
SOI & source of interest \nl
TDRSS & Tracking \& Data Relay Satellite System \nl
TJD & Truncated Julian Date (Julian Date - 2,440,000.5) \nl
WFC & Wide Field Camera \nl
\end{tabular}
\end{table}
\newpage
\section{BASIC FEATURES OF OCCULTATION TIMING}

The geometry and timing of Earth occultations of a celestial 
source are most simply understood for an idealized model where the orbit of 
the spacecraft is assumed to be circular, the precession of the orbital 
plane is ignored, and the oblateness of the Earth is ignored.

The basic geometry is shown in Fig. B1. We adopt a coordinate system in 
which the spacecraft orbit in in the x-y plane. The source is located at an
azimuthal angle $\phi$ from the x-axis, and an angle $\beta$ above the 
orbital plane. The spacecraft is located at radius $r_{sc}$ and 
azimuthal angle $\phi_{sc}$ which increases in time at a 
rate of $2\pi/P_{orbit}$. The angle $\theta $ between 
the direction to the source and the
direction to the Earth from the spacecraft is given by
\begin{equation}
\cos\theta = -cos(\phi-\phi_{sc})cos\beta~.
\label{EqB1}
\end{equation}
As shown in Fig. B2, this angle is related to $h$, the minimum height of
the line of sight above the surface of the Earth, by
\begin{equation}
r_e+h = r_{sc}\sin\theta 
\label{EqB2}
\end{equation}
where $r_e$ is the Earth radius.
In the middle of an occultation step we define 
\begin{equation}
\theta = \theta_{occ} = \sin^{-1}([r_e+h_{occ}]/r_{sc}) 
\label{EqB3}
\end{equation}
where $h_{occ}$ is
the altitude for 50\% transmission. For a spacecraft altitude of 500 km
and an occultation altitude of 70 km, $\theta_{occ} = 69.6^\circ$. 
If the source is too far above or below the orbital plane,  
$|\beta| > \theta_{occ}$, the source is always visible and no occultations
are seen. 
Earth spans  
an opening angle of $\sim$140$^{\circ}$, or about 30\% of the sky for
a spacecraft altitude of 500 km.   The oblateness of the Earth causes the
CGRO orbit to precess with a cycle of about 53 days.
Hence, during the precession cycle, if the angle in which a  vector in the
direction a source  
makes with the orbital plane ($\beta$) exceeds 69.6$^{\circ}$
occultations will cease.  The duration of this gap  in occultation coverage
is

\begin{equation}
\tau_{gap} = 2{{P_{precession}} \over {2 \pi}}
      \cos^{-1} \left({{\sin\theta_{occ}-\cos i \sin|\delta|} \over
                       {\sin i \cos \delta}}\right),
\end{equation}
where $\delta$ is the declination of the source, $P_{precession}$ is
the precession period, $\theta$$_{occ}$ is the angle between
the geocenter and the Earth's limb at 50\% transmission as seen from the
spacecraft, and $i$ is the inclination of the spacecraft orbit.
For a 500 km altitude orbit with $i$ = 28.4$^{\circ}$
$P_{precession}$ = 53.4 days.  We have then
\begin{equation}
\tau_{gap} = \Delta\Omega (6.74^{\circ}~day^{-1})^{-1}
\end{equation}
where
\begin{equation}
\Delta\Omega = 2\cos^{-1}{1.97\sec\delta-1.85\tan\delta}
\end{equation}
These time gaps occur for sources with declinations in the range
\begin{equation}
    \theta_{occ}-i < |\delta| < \pi-i-\theta_{occ}
\end{equation}
or $\pm$(41.2$^{\circ}$to 82$^{\circ}$).

Otherwise, a rise and set are seen each orbit at the spacecraft 
azimuth angles
\begin{eqnarray}
\phi^r_{sc} & = & \phi-\cos^{-1}(-\cos\theta_{occ}/\cos\beta) \\
\label{EqB4}
\phi^s_{sc} & = & \phi+\cos^{-1}(-\cos\theta_{occ}/\cos\beta) \nonumber
\label{EqB4B}
\end{eqnarray}
where the label $r$ is for source rise, $s$ for source
set. The duration of either occultation step is given by
\begin{equation} 
\Delta t_{occ} \approx 
\left({{P_{orbit}} \over {2\pi}}\right)
\left({{\Delta h} \over {r_{sc} \cos\theta_{occ}}}\right)
{{\sin\theta_{occ}} \over 
       {\sqrt{\cos^2\beta-\cos^2\theta_{occ}}}}~~,
\label{EqB5}
\end{equation}
where $\Delta h$ is the difference between the 90\% and 10\% transmission
altitudes.              

If we measure the times of a pair of rising and setting occultations, 
we can determine the location of the source. Fig. B3 shows 
the projection on the sky of the limb of the Earth at the time of rise and 
set for a source with $\phi = 0^\circ$ and $\beta = 50^\circ$. These rising
and setting limbs are the locus of directions $(\phi,\beta)$ 
that satisfy Eq.~(B8) with $\phi_{sc}$ fixed at its value at the
source rise or set. For a given pair of rising and setting occultations
there
are two possible source locations, one above and one below the orbital
plane.
This ambiguity must be resolved by using the direction sensitivity of
the detectors or other means.
If the two occultation times are measured with an accuracy of $\sigma_t$,
then the errors on the estimate for the source location can be shown to be
\begin{eqnarray}
\sigma_\phi &=& {{\sigma_t} \over {\sqrt{2}\Delta t_{occ}}}  
\left({{\Delta h} \over {r_{sc} \cos\theta_{occ}}}\right)   
{{\sin\theta_{occ}} \over 
       {\sqrt{\cos^2\beta-\cos^2\theta_{occ}}}}~~, \\
\sigma_\beta & = & {{\sigma_t} \over {\sqrt{2}\Delta t_{occ}}}  
\left({{\Delta h} \over {r_{sc} \cos\theta_{occ}}}\right)
{{\tan\theta_{occ}} \over {\tan\beta}}~~.    
\end{eqnarray}

\section{ACCURATE OCCULTATION CALCULATIONS}

Precise predictions of occultation times or calculations of the Earth limbs
at given times require the use of an accurate spacecraft ephemeris, and
an accurate model of the shape of the Earth. The Earth's surface is
approximately an oblate ellipsoid given by
\begin{equation}
x^2+y^2+(1-f)^{-2}z^2 = a^2
\label{EqC1}
\end{equation}
where $x$, $y$, and $z$ are geocentric Cartesian coordinates with the
z-axis aligned with the north pole, $f=1/298.257$ is the flattening 
factor of the Earth, and $a = 6378.136$\,km is the Earth's equatorial 
radius. Near the Earth's surface, constant atmospheric density surfaces 
can be approximated by ellipsoids of the same oblateness. If the spacecraft
is at position $R = (x_{sc},y_{sc},z_{sc})$, the source is in direction 
$\Omega = (\Omega_x,\Omega_y,\Omega_z)$,
and the height $h(s)$ above the surface of a point on the line of sight at a
distance $s$ from the spacecraft, then we have
\begin{equation}
(a+h(s))^2 = (x_{sc}+s\Omega_x)^2+(y_{sc}+s\Omega_y)^2
    +(1-f)^{-2}(z_{sc}+s\Omega_z)^2~~.
\label{EqC2}
\end{equation}
The minimum height, and its distance along the line of sight is given by
\begin{eqnarray}
h_{min} & = & \left[x^2_{sc}+y^2_{sc}+(1-f)^{-2}z^2_{sc}
 -{{(x_{sc}\Omega_x+y_{sc}\Omega_y+(1-f)^{-2}z_{sc}\Omega_z)^2} 
\over {\Omega^2_x+\Omega^2_y+(1-f)^{-2}\Omega^2_z}}\right]^{\onehalf}-a~~,
\\
s_{min} & = &  -{{x_{sc}\Omega_x+y_{sc}\Omega_y+(1-f)^{-2}z_{sc}\Omega_z}
\over {\Omega^2_x+\Omega^2_y+(1-f)^{-2}\Omega^2_z}}~~.
\end{eqnarray}
If $s_{min}$ is negative, then the minimum occurs in the direction from the
spacecraft away from
the source, and the source is visible, since that spacecraft is above
any significant atmosphere.

To accurately calculate the projection of the Earth's limb on the sky at a
given time, we take advantage of a linear transformation that maps the
oblate
Earth into a sphere. For a vector ${\bf X}$ we set
\begin{equation}
x = X_x,~~~~ y =X_y,~~{\rm and}~~z = (1-f)~X_z~~.
\end{equation}
Then Eq.~(\ref{EqC1}) reduces to $|{\bf X}|^2 = a^2$. The transformed
geometry of the occultation is like that discussed in Appendix A. Given the
spacecraft position $R$ and velocity $V$, we compute a set of orthonormal 
(in the transformed space) basis vectors:
\begin{eqnarray}
\mathbf{e}_1 & = & \mathbf{R}/|\mathbf{R}|  \\
\mathbf{e}_3 & = & \mathbf{R}\times\mathbf{V}/|\mathbf{R}\times\mathbf{V}|
\\
\mathbf{e}_2 & = & \mathbf{e}_3 \times \mathbf{e}_1
\end{eqnarray}
Then with $\theta_{occ} = \sin^{-1}([a+h_{occ}]/|\mathbf{R}|)$ we compute
the
direction vectors
\begin{equation}
\mathbf{\Omega}(\psi) = 
-\mathbf{e}_1 \cos\theta_{occ}+\mathbf{e}_2\sin\theta_{occ}\cos{\psi} 
                             +\mathbf{e}_3\sin\theta_{occ}\sin{\psi}
\end{equation} 
where the parameter $\psi$ ranges from $-\pi/2 < \psi < \pi/2$ for the
rising
limb, and $\pi/2 < \psi < 3\pi/2$ for the setting limb. After transforming
the direction vectors back to our original coordinate system we then
compute right ascensions $\alpha(\psi)$ and declinations $\delta(\psi)$ 
along the limb of the earth:
\begin{eqnarray}
\alpha(\psi) &=& \tan^{-1}(\mathbf{\Omega}_2(\psi)/\mathbf{\Omega}_1(\psi)) \\
\delta(\psi) &=& \tan^{-1}((1-f)\mathbf{\Omega}_3(\psi)
[\mathbf{\Omega}^2_1(\psi)+\mathbf{\Omega}^2_2(\psi)]^{-\onehalf}~).
\end{eqnarray}

The precession of the plane of the spacecraft orbit causes the orientation
of
the rising or setting limb passing through a source to change cyclically
with
the precession period ($\sim$50 days for a near Earth orbit with 23$^\circ$ 
inclination). If limbs are plotted for a new source for a number of days,
the
precession allows the location ambiguity discussed in appendix B to be
resolved.

\end{appendix}

\newpage
\section*{FIGURE CAPTIONS}

\figcaption{(a) The Compton Gamma Ray Observatory showing placement of the
Burst
and Transient Source Experiment detector modules. (b) Major components
of the BATSE detector modules (one of eight).
\label{fig:cgro.ps}}

\figcaption{DISCLA data (1.024 s resolution) in low and high energy bands 
at times in which the viewing
directions toward the Crab Nebula and the binary system A 0535+26 were close
to
face-on in a large area detector. To make the occultation steps more clearly
seen, zero suppression has been applied to the vertical
axes. (a) Approximately 7000 seconds of data in the 20-50 keV band,
where the background is dominated by
diffuse sky flux and earth shadowing. 
Crab occultation steps can be seen at
$\sim$550s (rise), 6050s (rise), and 4000s (set). (b) the same detector
in the 100-300 keV band, where the background is dominated by variations in
the
flux of secondary cosmic ray flux modulated by the
local magnetic field of the Earth, and (c) 20-50 keV band data containing a 
rise and a set
from occultation of the 
Be star/X-ray pulsar
system A 0535+26.  Data are from a giant outburst in 1994 Jan-Mar.
Individual
pulses (period 110 s) can clearly be seen between the rising and setting
features. Gaps in the data coverage result from either filtering or
temporary
loss of telemetry from CGRO to TDRSS (NASA Tracking and Data Relay
Satellite System).
\label{fig:discla.ps}}

\figcaption{Earth occultation step features for sources shown with a fit to
a quadratic model plus source terms modeled using attenuation by the 
atmosphere, where the fit assumes the background is continuous
before and after the step, and a linear fit, with independent slopes on
either side of the step. All fits are in CONT channel 4 (50-70 keV),
(a) and (b) show Crab steps and (c) and (d) show steps from the
transient black hole candidate GRO J0422+32 with Crab steps within the
four minute fitting window.  The linear model gives similar results
for the size of the step to the quadratic model except in the case of
(c) where the presence of the Crab step induces a systematic
error in the measurement of the GRO J0422+32 step. Vertical dotted lines
represent the computed occultation time for 100 keV photons at 50\%
transmission using the method described in Appendix B.
\label{fig:crab_step1.ps}}

\figcaption{Examples of count spectra and model fit residuals obtained from
CONT data for
the sources (a) Crab Nebula supernova remnant for TJDs 9783-9797 (b) the
black
hole candidate Cygnus X-1 for TJDs 10427-10434 in its low (hard) state
(c) the transient black
hole candidate GRO J1655-40 for TJDs 10322-10332 in its high or very high
state and
(d) and the neutron star high mass binary Vela X-1 for TJDs 10413-10420. The
dashed histograms represent the best fitting photon model folded
through the detector response.  Each observation includes two or more
LADs with angles between the source and the detector normal vector of
less than 60$^{\circ}$.  Residuals (measured counts - model counts)
in sigmas are shown in the lower frame.  The models and best fit parameters 
are given in the caption of Fig. 5. 
\label{fig:crab_9783_9797_cr.ps}}
\figcaption{Examples of spectra in photon space derived from the count spectra
of Fig. 4 for the
(a) Crab Nebula supernova remnant for TJDs 9783-9797 (b) the
black
hole candidate Cygnus X-1 for TJDs 10427-10434 in its low (hard) state
(c) the transient black
hole candidate GRO J1655-40 for TJDs 10322-10332 in its high or very high
state and
(d) and the neutron star high mass binary Vela X-1 for TJDs 10413-10420.  
All data and model results are in units of
photons cm$^{-2}$s$^{-1}$keV$^{-1}$. For (a) a broken power law was used
with a goodness of fit 49.85 for 38 d.o.f. and parameters: norm =
(3.46$\pm$0.023)$\times$10$^{-3}$@45 keV, $\alpha_{1}$=-2.08$\pm$0.015,
break energy = 136$\pm$15, $\alpha_{2}$=-2.43$\pm$0.056. For (b) the
Sunyaev-Titarchuk Comptonization model was used (Sunyaev \& Titarchuk
1980) with a goodness of fit 90.55 for 39 d.o.f. and parameters: norm
= (2.33$\pm$0.16)$\times$10$^{-4}$, kT = 49.4$\pm$1.2 keV and the optical
depth $\tau$=2.65$\pm$0.067 for a spherical plasma.  For (c) a single power
law was used with a goodness of fit of 59.53 per 54 d.o.f. and parameters
norm = (2.09$\pm$0.053)$\times$10$^{-4}$@100 keV, $\alpha$=-2.70$\pm$0.036.
For (d) an optically thin thermal bremsstrahlung model was used with
a goodness of fit of 11.6 per 16 d.o.f. and parameters norm = (5.31$\pm$2.2)
$\times$10$^{-6}$@100 keV, kT = 14.6$\pm$1.4.
\label{fig:crab_9783-9797.ps}}
\figcaption{(a) Examples of multi-year intensity histories 
(1991 April- 1998 July) for four persistent sources (from top to bottom)
the Crab supernova remnant (40-150 keV), the high mass binary pulsar
Vela X-1 (20-50 keV), the black hole candidate GX 339-4 (20-100 keV),
and the radio galaxy Centaurus A (20-200 keV) obtained with the Earth 
occultation
technique.  Each data point represents an average of occultation steps
obtained
for that day or several days.  (b) intensity histories for four transient
sources (from top to bottom) GRO J0422+32 (X-Ray Nova Persei 1992)
(40-150 keV), the high mass binary 2S 1417-624 (20-50 keV), GRO J1655-40
(X-Ray Nova Scorpii 1994) (20-200 keV) and GRO J1719-24 = GRS 1716-249
(X-Ray Nova Ophiuchi 1993) (20-100 keV).
\label{fig:persistent_sources_p1.ps}}

\figcaption{Sensitivity (3-$\sigma$) for a two-week observation of a source
averaging 16 occultations a day with contributions from two LADs, 
as a function of energy.  Energy bins correspond to CONT channel boundaries.
The Crab Nebula total emission spectrum measured with HEAO-A4
(Eq.~(\ref{eqn:S_Crab}))
(Jung 1989) is shown for comparison.
\label{fig:sens1.ps}}

\figcaption{Sensitivity curves (3-$\sigma$) for a two-week observation 
with the LADs as a function of 
angle from the 
normal vector of the LAD entrance window for combinations of one
(continuous curves), two (lower rows of diamonds), 
three and four LADs (higher rows of diamonds).  
Three different energy channels are shown.
\label{fig:sens2.ps}}

\figcaption{Sky regions near the (a) Crab Nebula and (b) the black hole
candidate
Cyg X-1.  Superimposed are shaded regions which represent the portion of the
sky 
subject to Earth occultation when the Earth's limb crosses the Crab or Cyg
X-1.   
Other sources which are routinely monitored with
BATSE are also shown. See text for a 
more detailed explanation.
\label{fig:fig8a.ps}}

\figcaption{(a) Angle between the closest approach points of a 
given
source location, here GRO J0422+32, with the rising (top frame) and setting 
(middle frame) limbs of the Crab Nebula as a function of time for the period
TJD 9800-10000.  The corresponding light curve of GRO J0422+32 for the
same period is also shown (bottom frame).  Note that the
closest approach of the setting limb of the Crab
to GRO J0422+32 is about 3$^{\circ}$. 
(b) Same type of plot for the source 4U 0614+091.  Note
that setting limb of the Crab periodically crosses the location 
of 4U 0614+091, 
when occultations occur at the same time for both sources. 
\label{fig:groj0422_limbs.ps}}

\figcaption{Sky grid, consisting of 162 points, used for determination of
location-dependent
systematic error.
\label{fig:grid.ps}}

\figcaption{Centroids in photons cm$^{-2}$s$^{-1}$ and width (in $\sigma$s)
of the 
light curves for the grid locations shown in Fig. 10 as a function of
galactic longitude.
\label{fig:avgsys.ps}}

\figcaption{Crab Nebula count rates in CONT channel 3 (40-50 keV) for
LADs 0, 1, 2, and 3 as a function
of the angle from the detector normal.  Data are binned in 10-degree
steps and weighted by the measured errors.
In the four top frames, the dotted curves represent the
original detector response function folded with the known Crab spectrum
(eq.~(\ref{eqn:S_Crab})) before the correction given in
eq.~(\ref{eqn:corr_fac})
was applied and, in the bottom four frames, the dotted curves represent
the response after the correction was applied.
The solid curve (same in the top and bottom frames) is an empirical fit to
the Crab
count data which reflects both the triangular response of the entrance
window and the cosine-like response due to the reduction in geometric
area toward the radiation source.  
\label{fig:crab_rates.ps}}

\figcaption{Histograms of the Crab Nebula daily average fluxes in
photons cm$^{-2}$s$^{-1}$ for the 20-100 keV band
before (upper frame) and after (lower frame) the response corrections in
eq.~(\ref{eqn:corr_fac}) were applied.  The histograms 
show the difference in the centroid and width of the distribution of flux 
measurements.
\label{fig:Crab_20-100keV_histograms.ps}}

\figcaption{Ilustration of the geometry of a lunar occultation of the Crab.
\label{fig:crab_moon_geometry.ps}}

\figcaption{Example of fits to BATSE CONT 40-50 keV data for LAD 7 for
lunar occultations of the Crab using an
extended source model for the Crab. 
\label{fig:channel3_lad7_extend.ps}}

\figcaption{Ratio of Crab lunar occultations to Crab Earth occultations
integrated over the 20-320 keV energy band. Each point represents the
ratio of the
average integrated intensity from lunar occultations to that of Earth
occultations for each detector and each day.
\label{fig:moon_earth_compare_extended.ps}}

\figcaption{Comparison of various instrument data for high energy
observations of the
Crab Nebula.  The shaded region represents a variation in the parameters
of the broken power law fit, and data from various instruments used by
permission (see text). 
\label{fig:crab_spec_mult_inst.ps}}

\figcaption{Results of the occultation step search.  The horizontal axis
represents the time elapsed from the start of the CGRO orbit and the
vertical
axis is the significance of the detection (size of the step in counts
divided by the uncertainty).  The number on the plot represents the number
of the LAD in which the most significant step of a detector
combination was found.
Times of bright source occultations are indicated on the plot, which 
correlate with the larger number of aligned bright steps detected by the
algorithm.
\label{fig:ocsearch_11628.ps}}  

\figcaption{A simulated image and associated occultation data in the time
domain
(after filtering) for two closely-spaced point sources.  Background data are
taken 
from one orbit on TJD 9573 and
one orbit on TJD 9575 (two rises and two sets) and combined with a simulated
source
signal.   This illustrates how a change in limb geometry can be used to
distinguish sources
within a few degrees of each other.
\label{fig:zh1-balanced_notext.ps}}

\figcaption{Examples of Earth occultation transform imaging for various
regions
along the galactic plane.
\label{fig:r-8389-8398-x1630-472-image.ps}}

\figcaption{Geometry of occulting source relative to the spacecraft orbit. 
\label{fig:fig_a1.ps}}

\figcaption{Geometry of the line of sight from the spacecraft to the source.
\label{fig:fig_a2.ps}}

\figcaption{The rising and setting limbs for a source at $\phi = 0^\circ$
and
$\beta = 50^\circ$. 
\label{fig:fig_a3.ps}}

\begin{deluxetable}{cccccc}
\tablenum{1}
\small
\tablecaption{Comparison of BATSE Earth Occultation Measurement Techniques
\label{tbl:occ_tech_compare}}
\tablewidth{0pt}
\tablehead{
\colhead{Occultation} & \colhead{Source} & \colhead{Datatype} & 
\colhead{Typical Energy} & \colhead{Sensitivity (3 $\sigma$-1 day)} &
\colhead{Source} \\
\tablevspace{-5pt}
\colhead{Technique} & \colhead{Location} & \colhead{} & \colhead{\& Channel}
& 
\colhead{(30-250 keV)} & \colhead{Localization} \\
\tablevspace{-5pt}
\colhead{} & \colhead{Assumption} & \colhead{} &
\colhead{Ranges\tablenotemark{a}} &
\colhead{$10^{-9}$ erg cm$^{-2}$ s$^{-1}$} & \colhead{Error(1
$\sigma$)\tablenotemark{b}}}
\startdata
Step Fit  & known & CONT & 20-1800 keV & 1.1 & $\gtrsim 0.2\arcdeg$ \nl 
\tablevspace{-5pt}
w/ atmos. model & \nodata & chans 1-14 & \nodata & \nodata \nl
Step Search & unknown\tablenotemark{c} & CONT & 30-250 keV & 4.6 & $\gtrsim
1\arcdeg$ \nl
\tablevspace{-5pt}
\nodata & \nodata & chans 2-8 & \nodata & \nodata \nl
Transform Imaging & unknown\tablenotemark{c} & DISCLA & 20-300 keV& 
1.1 & $\gtrsim 0.3\arcdeg$ \nl
\tablevspace{-5pt}
\nodata & \nodata & chans 1-3 & \nodata & \nodata \nl
\enddata
\tablenotetext{a}{Channel range is selectable.}
\tablenotetext{b}{Quoted error assumes optimal limb geometry for rise and
set, which should be regarded as a lower limit.}
\tablenotetext{c}{Detection of unknown weak sources may be improved by prior
subtraction
of known step-fit sources.}
\end{deluxetable}

\newpage
\begin{deluxetable}{cccc}
\tablenum{2}
\tablecaption{Times of lunar occultations of the Crab Nebula in BATSE CONT
data.
\label{tbl:moon_occ_times}}
\tablehead{
\colhead{Day (TJD)} & \colhead{Set Time (secs)} & \colhead{Rise Time (secs)}
& \colhead{LADs Viewing the Crab}}
\startdata
9103 & 10645 & 11150 & 0, 1\nl
9212 & 39061 & 39487 & 1, 3, 5, 7\nl
9239 & 74328 & 74543 & 1, 3, 5, 7 \nl
9376 & 20809 & 21036 & 5, 7\nl
9376 & 27098 & 27633 & 5, 7 \nl
\enddata
\end{deluxetable}

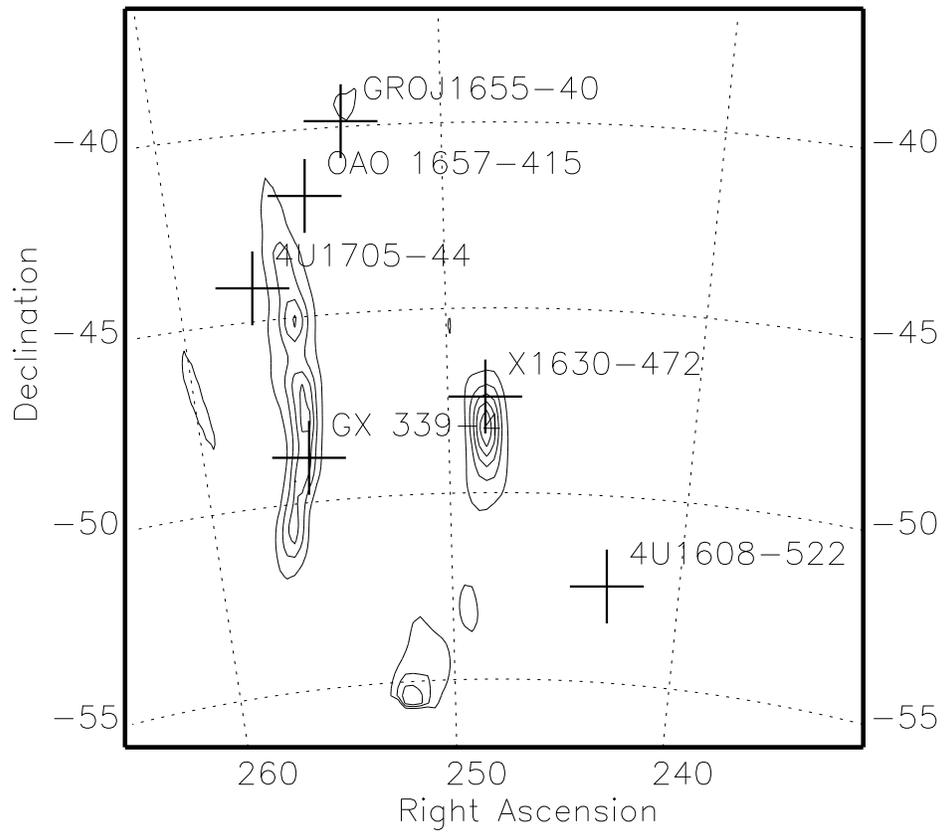
\end{document}